# Visualizing Energy Transfer Between Redox-Active Colloids

Subing Qu,[1,2,4] Zihao Ou,[1,2,4 †] Yavuz Savsatli,[1,2,4 †] Lehan Yao, [1,2,4] Yu Cao,[3,4] Elena C. Montoto,[3] Hao Yu, [4,5] Jingshu Hui,[1,2,3] Bo Li,[4,5] Julio A. N. T. Soares,[2] Lydia Kisley,[3,4] Brian Bailey,[1,2] Elizabeth A. Murphy,[3,4] Junsheng Liu,[1,2] Christopher M. Evans,[1,2,4] Charles M. Schroeder,[1,2,4,5] Joaquín Rodríguez-López,[2,3] Jeffrey S. Moore,[1,2,3,4] Qian Chen,[1,2,3,4,*] Paul V. Braun[1,2,3,4,*]

[1]Department of Materials Science and Engineering, [2]Materials Research Laboratory, [3]Department of Chemistry, [4]Beckman Institute for Advanced Science and Technology, and [5]Department of Chemical and Biomolecular Engineering, University of Illinois Urbana-Champaign, Urbana, Illinois 61801, United States

[†]These authors contributed equally to the work.

*To whom correspondence should be addressed. Email: qchen20@illinois.edu, pbraun@illinois.edu

## Abstract

**Redox-based electrical conduction in nonconjugated polymers[1] has been explored less than a decade[2], yet is already showing promise[3-5] as a new concept for electrical energy transport. Here using monolayers and sub-monolayers of touching micron-sized redox active colloids[6] (RAC) containing high densities of ethyl-viologen (EV) side groups, intercolloid redox-based electron transport was directly observed via fluorescence microscopy. This observation was enabled by the discovery that these RAC exhibit a highly non-linear electrofluorochromism[7] which can be quantitatively coupled to the colloid redox state. By evaluating the quasi-Fickian nature of the charge transfer (CT) kinetics[1], the apparent CT diffusion coefficient $D_{CT}$ was extracted. Along with addressing more fundamental questions regarding energy transport in colloidal materials, this first real-time real-space imaging of energy transport within monolayers of redox-active colloids may provide insights into energy transfer in flow batteries[8], and enable design of new forms of conductive polymers[9] for applications including organic electronics[10].**

**Introduction.** Electrically conductive conjugated polymers have been extensively studied[11] in contrast to nonconjugated conducting systems. While most nonconjugated polymers are not electrically conductive[2,12], in 2018, Boudouris showed that poly(4-glycidyloxy-2,2,6,6-tetramethylpiperidine-1-oxyl) (PTEO), a radical polymer glass, exhibited an electrical conductivity of 28 S/m over a distance of 600 nm[7] via a solid-state charge transfer (CT) mechanism. Unlike conjugated polymers, where conduction is exciton-based[13], and requires conjugation[11], which can be broken by defects, energy transport in nonconjugated systems is defect tolerant[5]. Required only is that redox groups be in proximity to form a solid-state percolating redox path or, as is the case for the non-conjugated ethyl-viologen-based redox-active colloids (RAC) serving as the basis of this study, have sufficient mobility to enable energy/electron exchanging contacts[6,14].

Based on our previous work[6,15], we knew these polymeric RAC were electroactive and their redox states could be reversibly accessed, which we attributed to self-exchange of electrons between the neighboring ethyl-viologen (EV) groups appended to polymer backbone. The net electron hopping is primarily driven by the concentration gradient of radical cations ($EV^{+\bullet}$) and dications ($EV^{2+}$) and exhibits a quasi-diffusional feature as proposed by Dahms and Ruff [16,17] (**Eq.1**),

$$D = D_{\mathrm{phys}} + D_{\mathrm{CT}} = D_{\mathrm{phys}} + \frac{k_{\mathrm{EX}}\delta^2 c}{6} \quad (1)$$

where the diffusion coefficient, $D$, contains both $D_{phys}$ and $D_{CT}$ terms. $D_{phys}$ represents the physical transport of polymer backbone (negligible in a crosslinked system such as the RAC studied here) and $D_{CT}$ is the CT diffusion coefficient. $D_{CT}$ is a function of $k_{EX}$, the self-exchange rate constant[18] for electron transfer between neighboring centers, c, the concentration of pendant group species and δ, the spacing between neighboring pendant groups. $D_{CT}$ is typically determined through electrochemical modelling[6] or via scanning electrochemical microscopy (SECM[19]) on single particle. Conventional electrical/electrochemical measurements cannot directly determine charge transport distances[5], therefore interparticle energy transport kinetics cannot be directly determined, motivating the development of methods to track the energetic state of assemblies of redox-active systems.

Enabled by the discovery the RAC are electrofluorochromic, we directly visualize electrical energy transport both between an underlying electrode and the RAC and within touching monolayers and sub-monolayers of these RAC. It is by observing and quantifying *in-situ* fluorescence waves moving order 10 µm across a colloidal layer only in contact on one side with the electrode that we extract the effective CT diffusion coefficient[20]. A key advantage of real-space imaging is the ability to extract full-field data, thus enabling determining how the structure of the 2D colloidal array impacts energy transfer.

**RAC Electrofluorochromism.** Under 488 nm excitation, we find the RAC fluorescence is highly dependent on the RAC redox state. Strong fluorescence is observed in the fully oxidized state ($EV^{2+}$) which is largely quenched when only 5-10% of the EV within the RAC are reduced to the $EV^{+\bullet}$ state (**Movie S1**, **Fig. 1g**). This system is the first example of visible light electrofluorochromic behavior in a viologen-based system[21], a significant expansion of the limited number of chemistries previously shown to be reversibly electrofluorochromic[22-27].

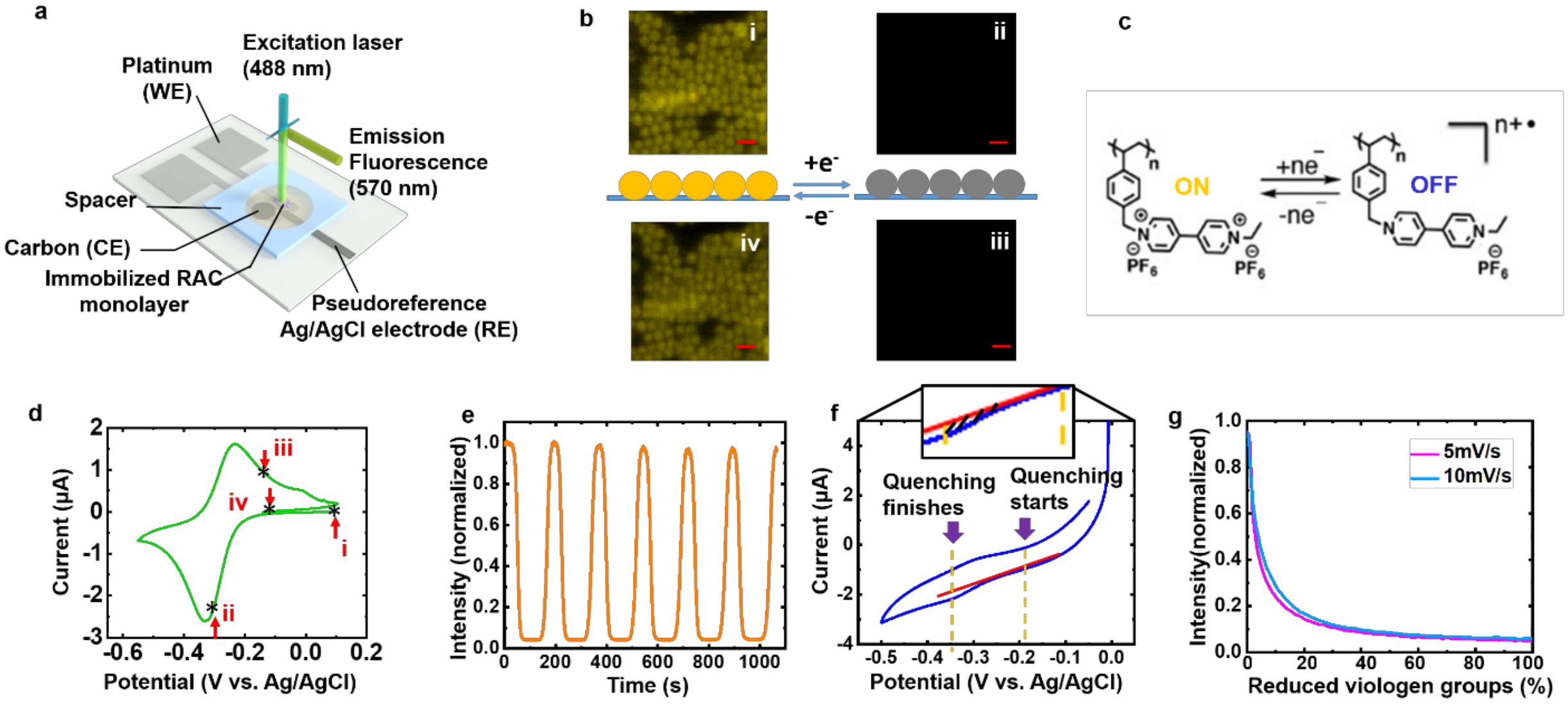

**Figure 1 | RAC Electrofluorochromism.** **a**, 3-electrode characterization cell schematic; **b**, Fluorescence imaging of RAC monolayer (scale bars are 2 µm) in oxidized (**i**, **iv**) and reduced (**ii**, **iii**) states; **c**, Reversible reduction and oxidation of the pendant EV group. The dication $EV^{2+}$ (oxidized) form is fluorescent, and the radical cation $EV^{+\bullet}$ (reduced) form is dark; **d**, Cyclic voltammetry (CV) on the RAC monolayer over the viologen first electron-transfer voltage range at 1 mV/s sweep rate. Fluorescence images presented in **b** are taken at the four indicated points (**i**, **ii**, **iii**, **iv**); **e**, Fluorescence intensity vs. time of a field of 166 colloids (**Supplementary Figure 18**) at a sweep rate of 5 mV/s between 0.0 and -0.5 V vs. Ag/AgCl; **f**, Tangent line drawing (indicatory) for extraction of Faradaic charges going into/out of RAC from the 1st round of 5 mV/s CV data in **Supplementary Figure 4**. Enclosed area (cross-hatched region of zoomed-in image between red tangent line and blue redox peak signatures) is the Faradaic charge; **g**, 1st cycle fluorescence intensity (normalized) vs. percent reduced EV groups at 5 and 10 mV/s sweep rates.

Using an airtight 3-electrode electrochemical cell compatible with fluorescence microscopy (**Fig. 1a**), 3-cycle CV and fluorescence imaging are concurrently performed on a solvent swollen RAC sub-monolayer at a scan rate of 1 mV/s (**Fig. 1b, Movie S1**). The difference between the reduction and oxidation peaks remains less than 100 mV (**Fig. 1d, Movie S1**), and fluorescence switching is reversible. Using a ferrocene calibrated pseudo-reference electrode (RE) (**M4 in SI**), the redox peaks are confirmed to be associated with cycling EV between $EV^{2+}$ and $EV^{+\bullet}$ states[14] (**Fig. 1c**). Four synchronized fluorescence images in the first CV cycle are presented in **Fig. 1b**. The RAC's fluorescence disappears during reduction and reappears during oxidation. As far as we can tell, the fluorescence of the entire volume of the colloid is switched.

The relationship between RAC fluorescence emission and state-of-charge is quantitatively determined by casting a few drops of a dilute (~0.04% w/w) solution of RAC onto the working electrode, and removing all the colloids outside the boundary of the electrode using an isopropanol-dipped wipe. The number of RAC (**Supplementary Figure 3**) participating in Faradaic reactions is precisely counted via a particle-counting method[28]. 6 CV cycles (**Supplementary Figure 4**) at 5 mV/s are performed on these 34,731 particles contacting the working electrode (WE) while simultaneously observing the fluorescence of 166 colloids (**Fig. 1e**). During the reduction sweep, Faradaic charge is extracted by the established tangent-line drawing method[20] (**Fig. 1f**) from a wave superimposed on a sloping baseline of capacitive current. Detailed steps of tangent-line fitting are shown in **SI** (**M8, M9**). Slower scan rates are not used, as the redox peaks become increasingly buried in the baseline under such conditions (**Supplementary Figure 14**), making quantitative analysis unreliable. The initiation of quenching is defined as when the total fluorescence intensity in the field of view decreased by 5%. Combining the Faradaic charge vs. time (**Supplementary Figure 6b**) and the luminescence intensity vs. time (**Fig. 1e)** enables determining the fluorescence vs. redox state (**Fig. 1g**) of the RAC. Only small differences are observed at 5 and 10 mV/s. Electrochemical data over the entire electrode is integrated to obtain a larger readable current signal, while imaging a much smaller number of particles to obtain single particle fluorescence data. Low-magnification experiments (**Movie S9**) show the fluorescence of the majority of colloids (~95%) increases and decreases in unison. We suspect the ~5% of colloids which do not follow the behavior of the others are simply in poor contact with the electrode. The working curve for the 1st cycle is selected because it reflects the original solution environment (e.g., electrolyte chemistry, double-layer structure, concentration gradient of varying species), and a minimum of potentially photobleached RAC.

$$\mathbf{I_o/I = 1+K_s[Q] \quad (2)}$$

Interestingly, fluorescence quenching is a strong function of redox state. Fluorescence decreases roughly linearly by ~80% as the first 13% (5 mV/s) to 16% (10 mV/s) of the EV groups are reduced (**Fig. 1g**), indicating 1 $EV^{+\bullet}$ quenches ~ 4 or 5 $EV^{2+}$ within a nearly fully oxidized RAC. **Supplementary Figure 16** repeats **Fig. 1g** for Cycle 2 and 3 at 5 mV/s. There is a reasonable overlap between subsequent cycles, however, since the coulombic efficiency is less than 100%, subsequent cycles do not start in the fully oxidized state. The fact that fluorescence quenching is a strong function with respect to EV reduction requires each $EV^{+\bullet}$ group to quench multiple $EV^{2+}$ groups. **Fig. 2a** is a Stern-Volmer[29] plot obtained by measuring the photoluminescence (PL) intensity at 570 nm of an acetonitrile suspension of oxidized RAC (containing a total of 10 mM EV) as chemically reduced ethyl-viologen monomer ($EV^{+\bullet}$) was added stepwise. We note the PL decreased significantly after only 0.08 mM of reduced monomer was added (**Supplementary Figure 8**). Assuming quenching follows a Stern-Volmer relationship (**Eq. 2**), $K_s$, the quenching constant, can be obtained through a linear fit and is found to be $5.34\times10^4$ $M^{-1}$. Note the fit did not go through the origin, which we speculate is due to an oxidant present in the solution such as $O_2$ which oxidizes some of the added $EV^{+\bullet}$. The data in **Fig. 2a** suggests one $EV^{+\bullet}$ monomer can quench emission from ~79 $EV^{2+}$ pendant groups on the polymer backbone (**M11 in SI**), and also suggests that the $EV^{+\bullet}$ monomer diffuses readily into the acetonitrile-swollen RAC, as otherwise significant fluorescence would remain even in the presence of $EV^{+\bullet}$.

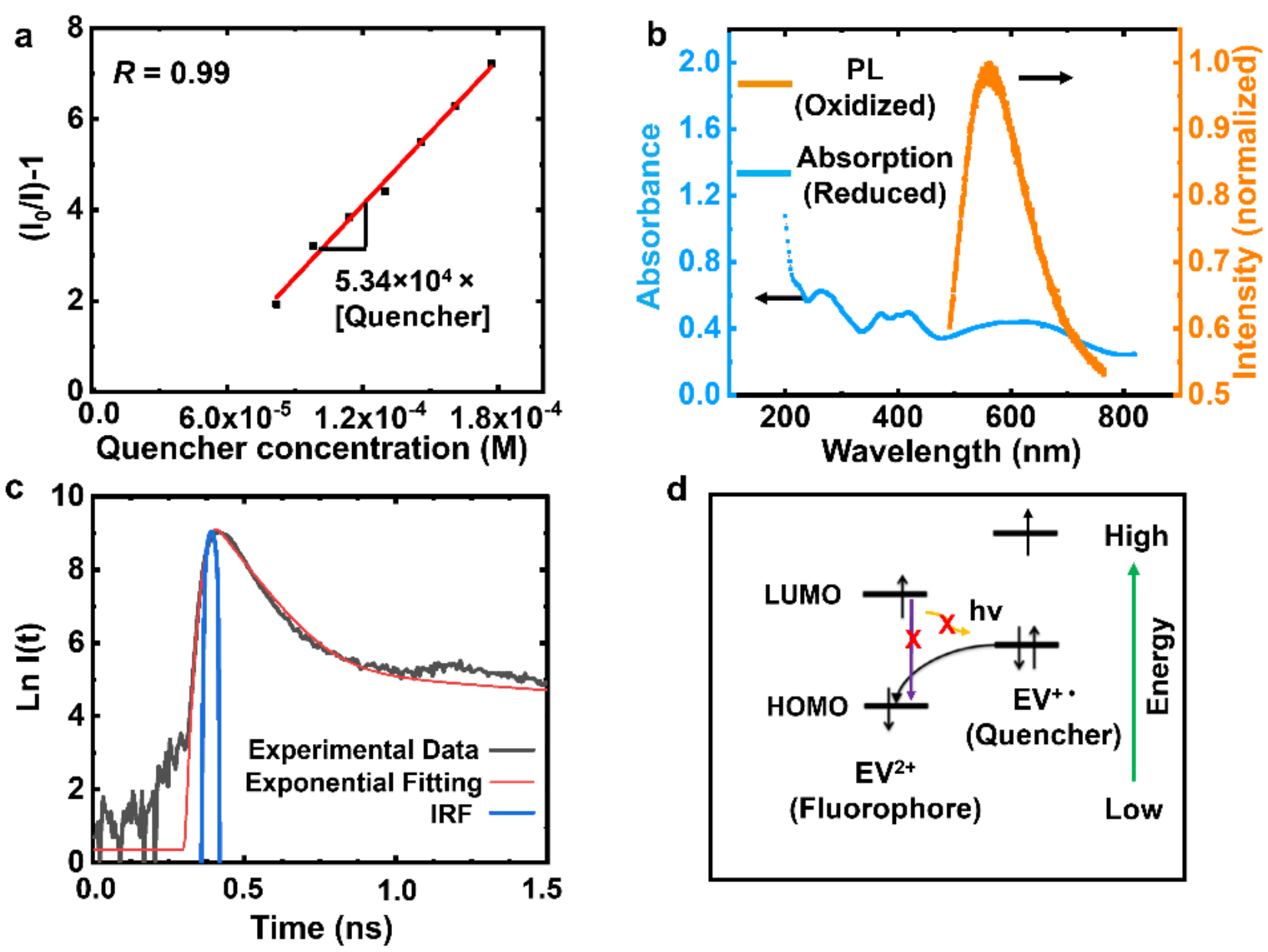

**Figure 2 | RAC quenching photophysics. a**, Stern-Volmer plot of reduced monomer ($EV^{+\bullet}$) quenching oxidized RAC dispersion (10 mM total viologen groups); **b**, Spectral superimposition of reduced RAC (0.1 mM) absorbance and normalized photoluminescence (PL) emission for oxidized RAC (under 488 nm excitation); **c**, Time-resolved photoluminescence measurement and single-exponential fit of fluorescence lifetime of oxidized RAC dispersion (0.1 M viologen groups), IRF=Instrument Response Function; **d**, Proposed electron state diagram for electron-transfer based quenching in RAC system.

We do not know the quenching mechanism, however, speculate quenching is due to the two most common mechanisms for fluorescence quenching[30], non-radiative energy transfer and electron transfer. Non-radiative energy transfer can occur either through Forster Resonance Energy Transfer (FRET) or a Dexter mechanism[7], both of which require significant spectral overlap between the donor and acceptor species. Based on the absorbance data in **Fig. 2b**, the reduced RAC has a weak absorbance peak around the oxidized RAC's emission peak wavelength (570 nm). Given the high concentration of EV groups in the swollen RAC (order one EV/$nm^3$), sufficient EV groups are within a conventional FRET distance of ~5 nm for a FRET quenching mechanism to operate. However, the low absorbance of the reduced RAC suggests this effect may not be the sole factor contributing to quenching. Electron transfer, see the proposed electron state diagram in **Fig. 2d**, may also play a role in quenching. Once an $EV^{2+}$ is excited, one of the paired electrons in $EV^{+\bullet}$ can transfer to the now available highest occupied molecular orbital (HOMO) of the excited $EV^{2+}$, inhibiting relaxation of the excited electron from the $EV^{2+}$ lowest unoccupied molecular orbital (LUMO), quenching the fluorophore. Time-resolved photoluminescence shows a fast fluorescence lifetime of ~0.1 ns (**Fig. 2c**), along with a much slower decay. 0.1 ns is many-fold shorter than the time[31] (inversion of $k_{EX}$ reported for viologen-based redox active polymer, which is ~$8.1\times10^6$ $s^{-1}$) needed to complete a single step electron self-exchange, indicating the non-linearity of RAC's electrofluorochromism is probably due to non-radiative energy transfer,but we cannot rule out other possibilities.

**Interparticle Energy Transport.** In the previous section, the RAC were in direct contact with an underlying electrode. Colloids in the near vicinity but not contacting the electrode also exhibit fluorescence switching as long as the colloids formed a percolating pathway to the electrode. Using this discovery, *in-situ* fluorescence imaging was used to visualize interparticle electron transport during electrochemical cycling. Assuming electron transport occurs via thermally-activated electron hopping as described by Dahms and Ruff[32,33], charge transport will occur between physically touching colloids (**Fig. 3a**) provided there is a concentration gradient. To quantify interparticle energy transport, the front edge of the fluorescence (for detailed methodology see **M15** in **SI**) is tracked as the electrode potential is switched between oxidizing and reducing (**Fig. 3c**). Note, colloids above the platinum electrode always appear brighter than those above insulating regions of the substrate (**Fig. 3b**) because of reflections from the electrode, and possibly metal-enhanced fluorescence[34]. By synchronizing the movies (**Movies S4-7**) with the electrochemical potential, the relationship between the electrode potential and redox state of the colloids both in direct contact, and connected via a percolating pathway with the electrode is determined. In **Fig. 3b**, the RAC are initially in the oxidized (fluorescent) state. At t=0 s, the electrode is switched to -0.6 V. The fluorescence of RAC on the electrode are significantly quenched within ~10 s. By ~100 s, the fluorescence front propagates past the edge of the electrode (denoted by the green dashed line in **Fig. 3b,** choosing criteria shown in **M15** in **SI**) into the percolating colloids on the glass. After about 800 s the front stops moving. When the electrode potential is switched to +0.2 V, the fluorescence front retreats towards the substrate with the RAC on the electrodes recovering last (**Fig. 3b**). The fluorescence front propagates multiple colloid diameters away from the electrode, suggesting interparticle electron transport. As evidence for the requirement for the RAC to be in physical contact for energy transfer to occur, in control experiments where isolated clusters of RAC were present near, but not touching the electrode, the fluorescence of these clusters does not change during electrochemical cycling (**Supplementary Figure 12 and Movie S8**), strong evidence that energy transport is not occurring via a solvent-mediated process.

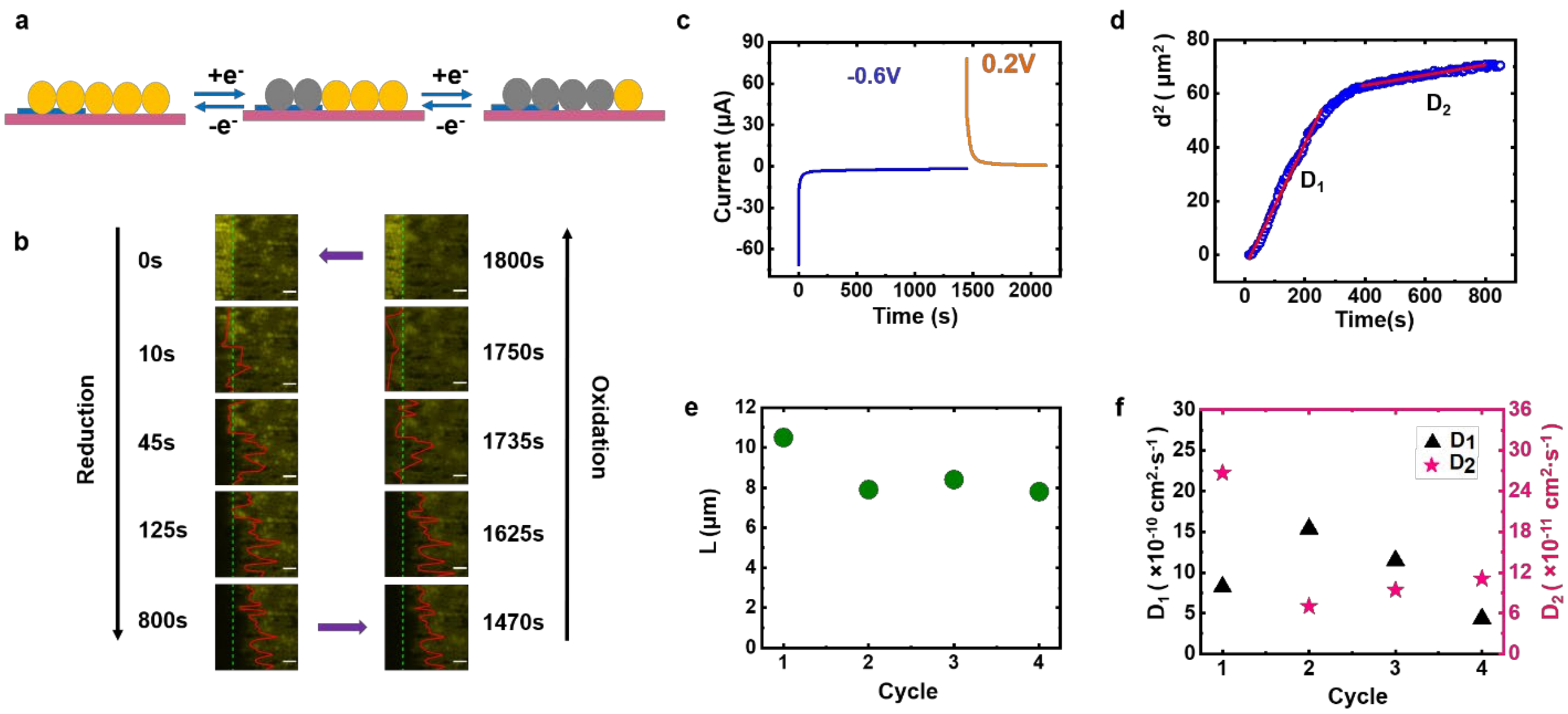

**Figure 3 | Inter-colloid electron diffusion. A**, Schematic illustrating interparticle lateral charge transport; **b**, Fluorescence images with fluorescence front tracking (cropped from **Movie S6**), during both reduction and oxidation. Scale bars are 5 µm; **c**, Chronoamperometry (CA) performed on a RAC monolayer, with reduction potential held at -0.6 V vs. Ag/AgCl and oxidation potential held at +0.2 V vs. Ag/AgCl; **d**, Tracking result from **Movie S6**: Lateral distance squared vs. time (extrapolated from **Supplementary Figure 17**), with least-squared linear fits of two sections of the data; **e**, Lateral quenched limit (L) tracked for 4 cycles of CA experiments in the reduced state collected from the same RAC region used to collect the data in **d**; **f**, Lateral charge-transport diffusion coefficient ($D_1$ and $D_2$) for 4 cycles performed in the same RAC region as in **d**, note the different y-axis scales.

The reduction cycle is rather easy to understand. The RAC in direct contact with the electrode are reduced first, and then subsequently RAC increasingly farther away from the electrode are reduced. The oxidation cycle is a bit more complex. Naively, one would assume RAC directly in contact with the electrode would switch first, and RAC away from the electrode last. However, that is not the case. The RAC furthest from the electrode switch first, and those in contact with the electrode last. It is important to remember that the RAC only fluoresce when almost fully oxidized (see **Fig. 1g**). What we believe is happening is that the switched colloids furthest from the electrode are only slightly reduced, and thus when the electrode is switched, it is those RAC that fluoresce first, and then the colloids above the electrode, which are the most fully reduced (oxidized colloids even further away from the electrode can also serve as a sink for electrons). During oxidation, we sometimes did observe RAC on the electrode recovering their fluorescence almost immediately. We assume these are the particles which are not percolated into the RAC network, and thus oxidize more rapidly than the RAC around them.

We noted that after ~800 s, the fluorescence front appears to stop ~8 μm away from the edge of the electrode (see tracking data in **Supplementary Figure 17**). This was repeated over multiple cycles except the fluorescence front appeared to stop ~10 μm from the edge of the electrode after the first cycle (**Fig. 3e** and **Movies S4-S7**). The data in **Fig. 3d** corresponds to Cycle 3 **(Movie S6)** in both **Figs. 3e, f**. We can only speculate why the front stops moving ~10 μm from the electrode edge, however several possibilities include electron hopping kinetics decay due to increasing energy barrier over certain distance of cumulative defects (e.g., cracks, poor percolation of colloids, or misalignment of viologen pendant groups), and Joule heating of the colloidal wire. One possibility is that the colloids are more tightly percolated near the electrode, however, optical imaging indicates the degree of percolation was uniform throughout.

$$d^2 = 2D_{CT}t \qquad (3)$$

As already mentioned, we can only speculate why fluorescence quenching stops ~10 μm from the electrode, however, it was still possible to quantify charge transfer kinetics in the RAC monolayer. By plotting the lateral propagation distance squared ($d^2$) against time (t) (**Fig. 3d**), we observe two distinct Fickian regions[20] (**Eq. 3**) with least-squared linear fits of 0.229 and 0.0189 $\mu m^2/s$ for their slopes. Since the slope is twice the numerical value of the effective CT diffusion coefficient $D_{CT}$, we calculated $D_1$ to be $1.15\times10^{-9}$ $cm^2/s$, and $D_2$ to be $9.45\times10^{-11}$ $cm^2/s$. These values are in the same order of magnitude as reported for other viologen-containing polymers[6,19,35] ($10^{-11}$ to $10^{-10}$ $cm^2/s$), giving us some confidence that our physical interpretation of the data is reasonable. $D_1$ and $D_2$ over 4 cycles are shown in **Fig. 3f**. We note $D_{CT}$ here is a convolution of both intra-colloid and inter-colloid electron hopping and we do not know which one of these two processes is rate-determining. While decoupling intra- and inter-colloid resistance is challenging, advanced imaging techniques with higher spatial and temporal resolution (e.g., super-resolution microscopy[36,37]) may enable decoupling intra- and inter-colloid electron transport.

**Conclusion**. The most important coupled findings are 1) observation of highly non-linear RAC electrofluorochromism, 2) demonstration of intercolloid energy transport over multiple colloids, 3) development of a simple method to determine $D_{CT}$ from real-space imaging. By linking charge state with fluorescence intensity and using fluorescence microscopy to visualize electron/energy transport in a densely-packed, well-percolated RAC monolayer, quantification of the kinetics of electron transport through colloid-metal and colloid-colloid contacts is enabled. We note, the observed ~10 μm lateral charge diffusion in nonconjugated polymers is an order of magnitude greater than previously observed in other radical polymers[5,6,11,13], possibly due to the contributory effect of solvent molecules' Brownian dynamics. We suggest the direct visualization of energy transfer in these redox-active colloids will be extendable to the suspension state, enabling study of pairwise interactions, addressing the key question of how two soft colloids at different energetic states interact upon collision.

## Methods

Methods, including statements of data availability and any associated accession codes and references, are available at Supplementary Information (**SI**).

## Acknowledgements

This work was supported by the U.S. Department of Energy, Office of Basic Energy Sciences, Division of Materials Sciences and Engineering under Awards No. DE-FG02-07ER46471 and DE-SC0020858.

# Supplementary Information

## Visualizing Energy Transfer Between Redox-Active Colloids

## Authors

Subing Qu,[1,2,4] Zihao Ou,[1,2,4 †] Yavuz Savsatli,[1,2,4 †] Lehan Yao, [1,2,4] Yu Cao,[3,4] Elena C. Montoto,[3] Hao Yu, [4,5] Jingshu Hui,[1,2,3] Bo Li,[4,5] Julio A. N. T. Soares,[2] Lydia Kisley,[3,4] Brian Bailey,[1,2] Elizabeth A. Murphy,[3,4] Junsheng Liu,[1,2] Christopher M. Evans,[1,2,4] Charles M. Schroeder,[1,2,4,5] Joaquín Rodríguez-López,[2,3] Jeffrey S. Moore,[1,2,3,4] Qian Chen,[1,2,3,4,*] Paul V. Braun[1,2,3,4,*]

[1]Department of Materials Science and Engineering, [2]Materials Research Laboratory, [3]Department of Chemistry, and [4]Beckman Institute for Advanced Science and Technology, [5]Department of Chemical and Biomolecular Engineering, University of Illinois Urbana-Champaign, Urbana, Illinois 61801, United States

[†]These authors contributed equally to the work.

*To whom correspondence should be addressed. Email: qchen20@illinois.edu, pbraun@illinois.edu

## Materials and Methods

### M1. Synthesis of RACs

The RACs were synthesized following our previously published procedure.[1]

### M2. Photolithography procedures for device fabrication and electron-beam deposition

The 3-electrode device for *in-situ* imaging is fabricated in the cleanroom as follows. The electrodes are patterned onto glass slides with dimension of 75 mm × 50 mm. The glass slides are first cleaned by acetone, isopropanol (IPA), deionized water and IPA in sequence and blown dry with nitrogen gas. The glass slide is then dehydrated on a hot plate at 110 °C for 5 minutes. Photoresist (AZ5214E) is then spincoated onto the glass slide with spin speed of 3000 rpm for 30 s. The glass slides are then baked on a hot plate at 110 °C for 1 min and then exposed under UV light for 9 s using a photomask and developed using AZ917 for 55 s. The glass slides are immediately immersed in a DI water bath for a few seconds and then blown with nitrogen gas. After development, the inversed image of the electrode is already visible to eye and the devices are baked on a hotplate at 110 °C for 3 minutes. After photolithography, 50 nm Pt is deposited using electron-beam evaporation on top of 5 nm Cr as adhesion layer. Subsequently, the device is sonicated in acetone for 5 minutes to eliminate all the materials except for the metallic electrodes.

### M3. Preparation of counter electrode (CE) and pseudoreference electrode (RE)

To prepare the counter electrode, carbon black, super P (99% from Alfa Aesar) and poly (vinylidene fluoride) (PVDF) (from Sigma Aldrich) are mixed with a few drops of N-Methylpyrrolidone (NMP) (99% from Sigma Aldrich) using a pestle and mortar for 20 minutes. Then a small piece of slurry is cast on one of the Pt electrodes and the substrate is heated at 205°C for 45 minutes to ensure NMP removal. After letting the device cool down in ambient air for 10 minutes, Ag/AgCl (65:35 from Creative Materials) is deposited as a paste on the device as the reference electrode and cured 30 minutes at 175 °C.

### M4. Pseudoreference electrode potential calibration

On the fabricated device with the pasted pseudoreference Ag/AgCl electrode (RE), 20 mV/s cyclic voltammetry (CV) experiments are run on a 10 mM ferrocene solution with 0.1 M $LiBF_4$ as supporting electrolyte in acetonitrile. Based on peak voltage values, ferrocene oxidation takes place at ~ 0.59 V vs. RE, which according to literature[2] value signatures the ethyl viologen first reduction taking place at ~ -0.22 V vs. RE. This aligns with **Fig. 1d** in the main text.

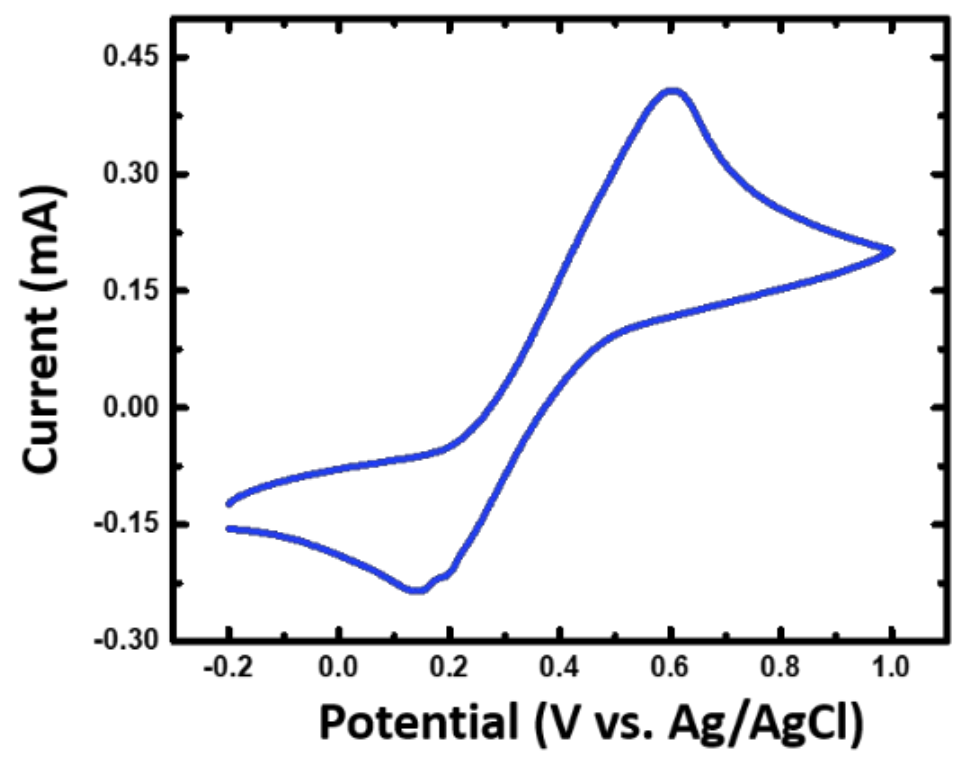

**Supplementary Figure 1. Ferrocene cyclic voltammetry (20 mV/s) in the planar device used for fluorescence imaging.**

### M5. Fluorescence imaging of the redox active colloids (RAC)

The fluorescence imaging experiments are carried out on a ZEISS LSM7 LIVE system with a 20X

(EC Plan-Neofluar 20x/0.50 M27), 40X (Plan-Apochromat 40x/1.4 Oil DIC M27) or 100X (Plan-Apochromat 100x/1.40 Oil M27) objectives depending on different magnification requirements in different experiments. The wavelength of excitation laser is 488 nm and a filter is applied to block all the light with wavelength below 495 nm. The glass slide is mounted on the sample holder and connected to the portable potentiostat with copper tape. **Supplementary Figure 2d** shows a photo of the whole setup for *in-situ* study of colloidal fluorescence while conducting electrochemical cycling.

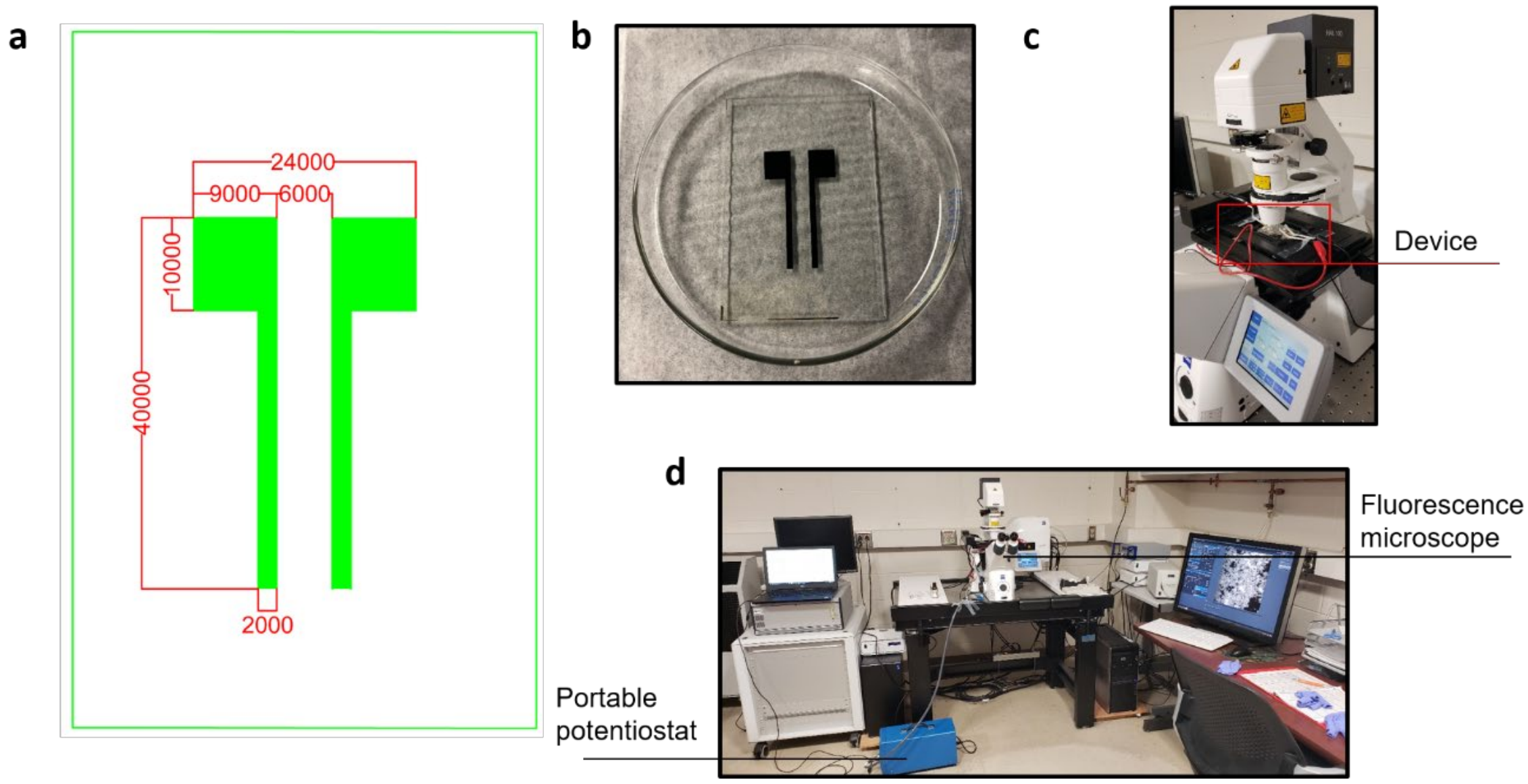

**Supplementary Figure 2. Device design, device photo and real experimental setup.** (a) Photomask design drawn in CAD. Dimensions in μm. (b) Fabricated Pt electrodes on the glass slide. (c) Device mounted on the imaging stage of fluorescence microscope. (d) Experimental set-up including fluorescence microscope, portable potentiostat, device and laptop.

## M6. Particle counting technique

Particle counting was conducted by a home-built MATLAB code suite. Different from the zoomed-in view shown in **Supplementary Figure 18**, a large-scale scanning of the whole sample area was conducted utilizing the "Tiling" module in the confocal microscope to keep single particle resolution, see **Supplementary Figure 13**. A gaussian filter was first applied onto the original image (built-in function imgaussfilter.m with standard deviation value of 0.01). Then centroid positions of each particle were tracked based on the peak intensity positions of each particle utilizing the codes developed for optical microscopy[3] (pkfnd.m and cntrd.m from the Ref. 3). Tracking results and zoomed-in views of a typical region are shown in **Supplementary Figure 3**. Red dots in **Supplementary Figure 3c** denote centroid positions of single colloid, whose intensity values were tracked for each snapshot to extract the temporal intensity evolution in **Fig. 1e** and **Supplementary Figures 6a, 15b**.

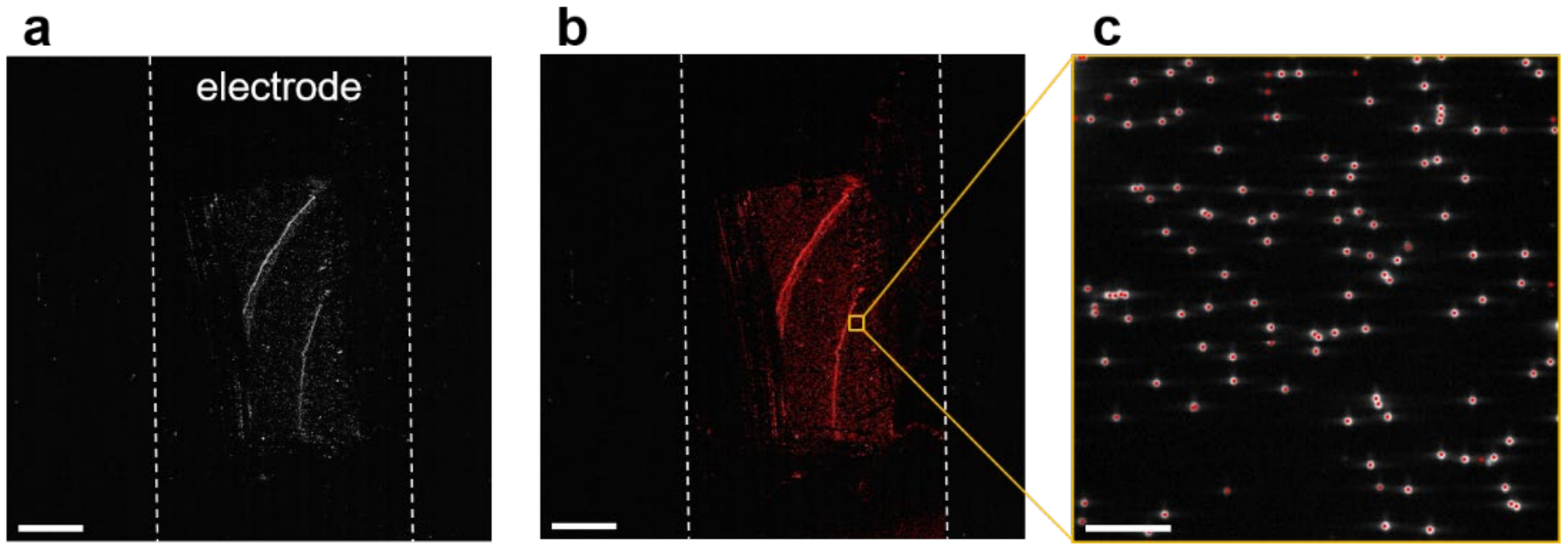

**Supplementary Figure 3. Tracking and counting the total number of RACs on the electrode.** (a) Raw fluorescence image of colloids on the electrode. The doted lines denote the boundary of the electrode. (b) Raw fluorescence image overlaid with the particles on electrode tracked shown as the red dots (34,731 colloids). (c) Zoomed-in view showing a typical region on the electrode with centers of the colloids highlighted by red dots. Scale bars: 500 µm in (a-b) and 20 µm in (c).

## M7. Number of ethyl viologen groups in a single RAC and on the entire electrode used in M6

$Number$ of pendant groups in a single RAC (Dry state) $= \frac{\rho_1 \times V \times N_A}{M_w}$ = 1.25 g/cm$^3$ × ($\frac{1}{6}\pi \times (9.5 \times 10^{-5} cm)^3$) × 6.022 ×10$^{23}$÷ 592 g/mol = 5.71×10$^{8}$

Total number of ethyl-viologen groups in 34731 particles = 5.71×10$^{8}$ × 34731≈ 1.98×10$^{13}$

- Dry state density of RAC was reported before[1].

## M8. Tangent line fitting[4] for extracting Faradaic charges in the cyclic voltammetry plot

The Faradaic charges which are used for redox reactions as opposed to resistive or capacitive charges (non-Faradaic) are drawn out from the CV plot displayed in the software EC-Lab. Demonstrated in **Supplementary Figure 5a**, by subtracting the real-time trapezoidal area enclosed with $i_1$, $i_2$, x axis and the tangent line from the integrated area of I vs. Potential (potential could be easily converted to time divided by scan rate), we can plot the $\mathbf{Q_{Faradaic}}$ over time as presented in **Supplementary Figure 6b**. Quenching start point is chosen when the total fluorescence in the field of view decays no smaller than 5%, and quenching ending point is chosen when the total fluorescence in the field of view remains no larger than 6%. See the example (Cycle 1 at 5mV/s) in **Supplementary Figure 5a**. In this example, the quenching start potential is ~ -0.19 V. We select 400 mV positive and 200 mV negative than this value of -0.19 V as the upper and lower bounds of linear fitting (61 data points in total, see inset in **Supplementary Figure 5b**). The same method to determine voltage windows is applied for all the other linear fittings in this paper (**Supplementary Figure 5c** for instance). The only difference is that for Cycles 2 & 3, since the electrochemistry environment has shifted, we see a background shift (upward and leftward) in CV as well as some pronounced shape change compared with Cycle 1. To ensure that result of tangent line fitting makes physical sense, we move the tangent point to 300 mV more negative (**Supplementary Figures 5d & e**). The number of data points selected (61) and approach of choosing upper and lower bounds remain.

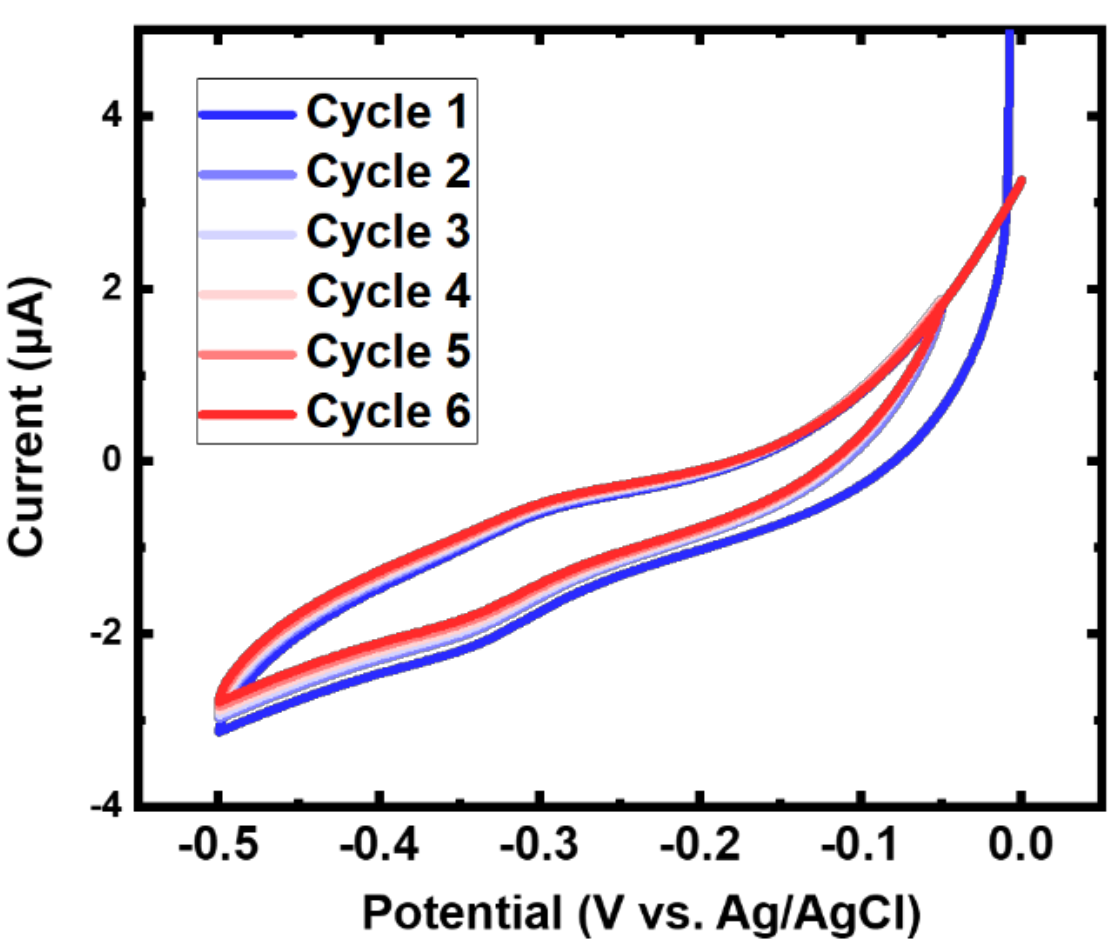

**Supplementary Figure 4. 6-round CV on discrete particles (total number: 34,731) at 5 mV/s.**

**a** **5mV/s_Cycle 1 (Fluorescence Quenching): 36.3s~66.0s**

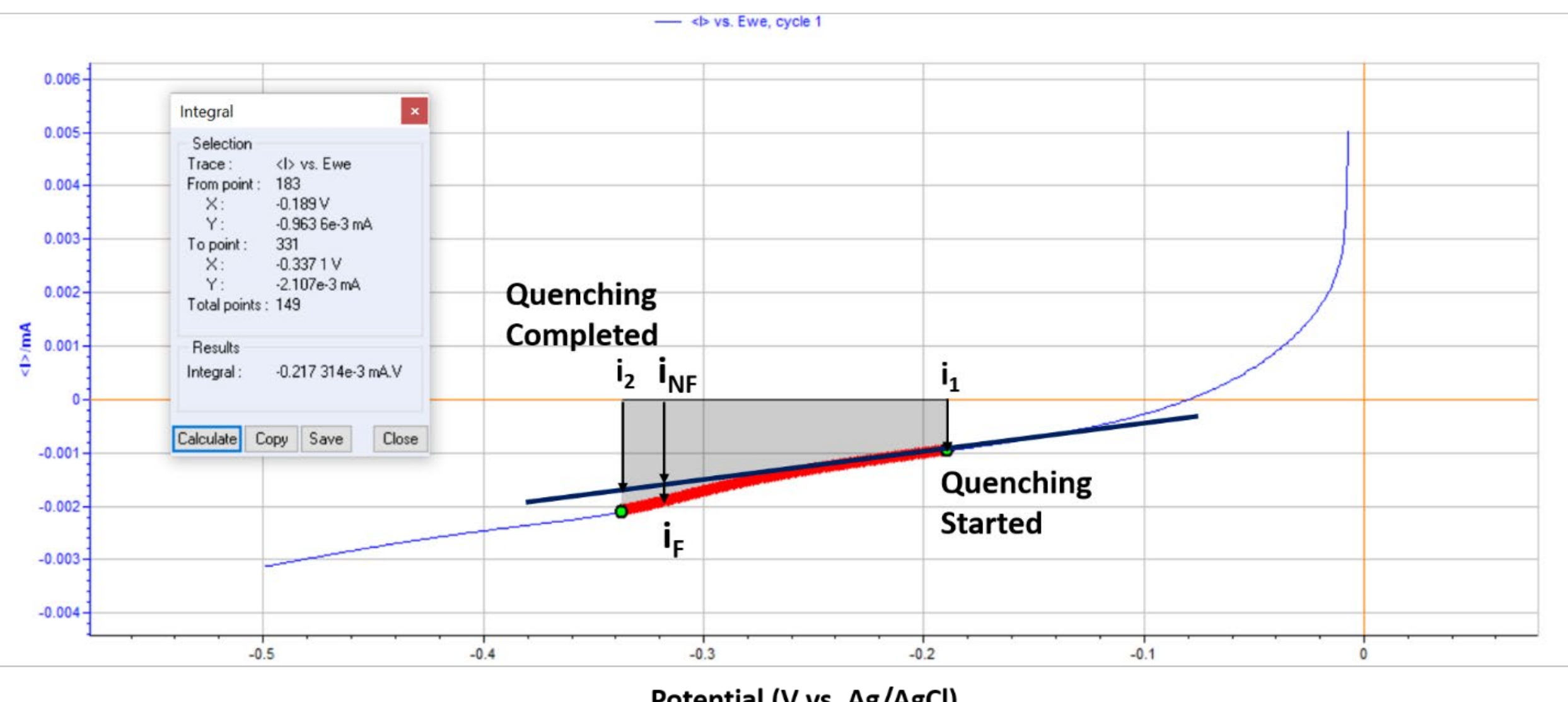

**b 5mV/s_Cycle 1 linear fitting**

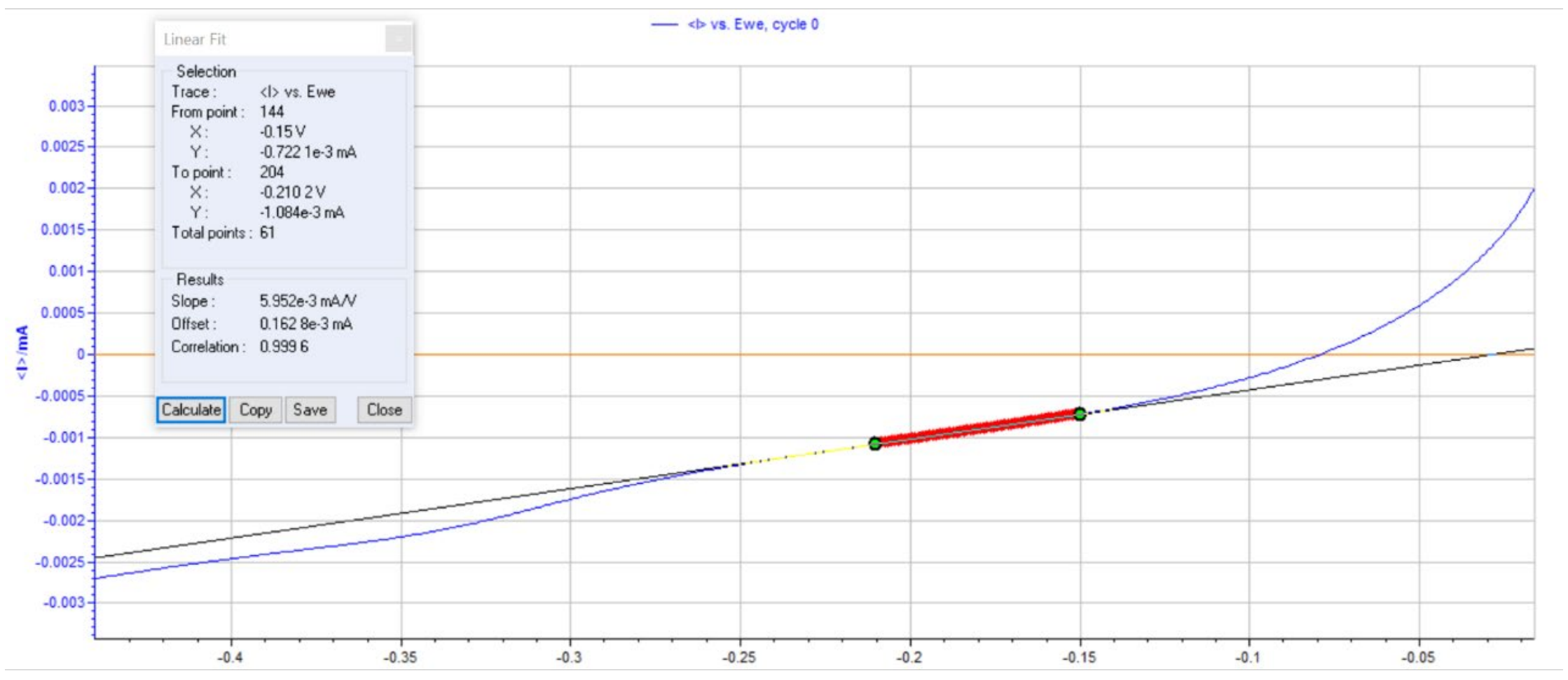

**Potential (V vs. Ag/AgCl)**

**c 10mV/s_Cycle 1 linear fitting**

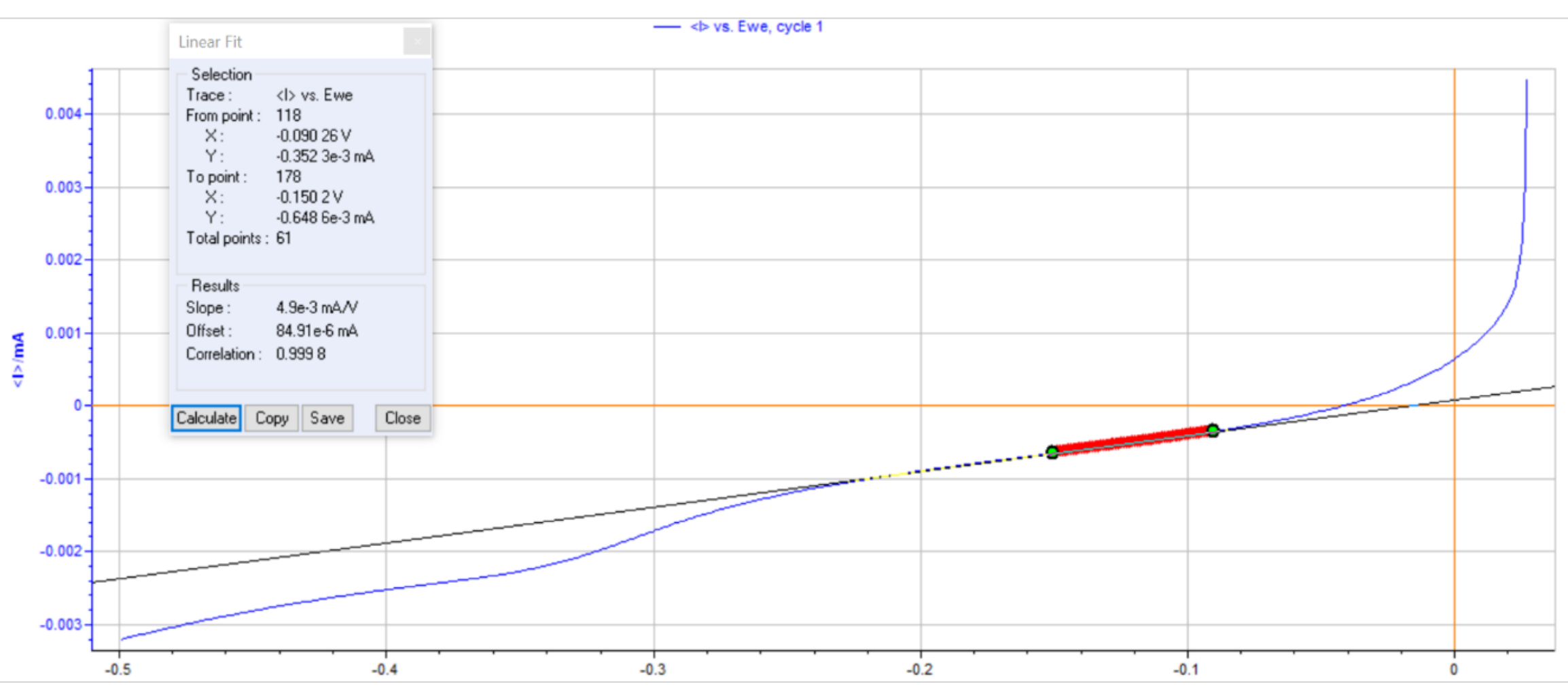

**Potential (V vs. Ag/AgCl)**

**d 5mV/s_Cycle 2 linear fitting**

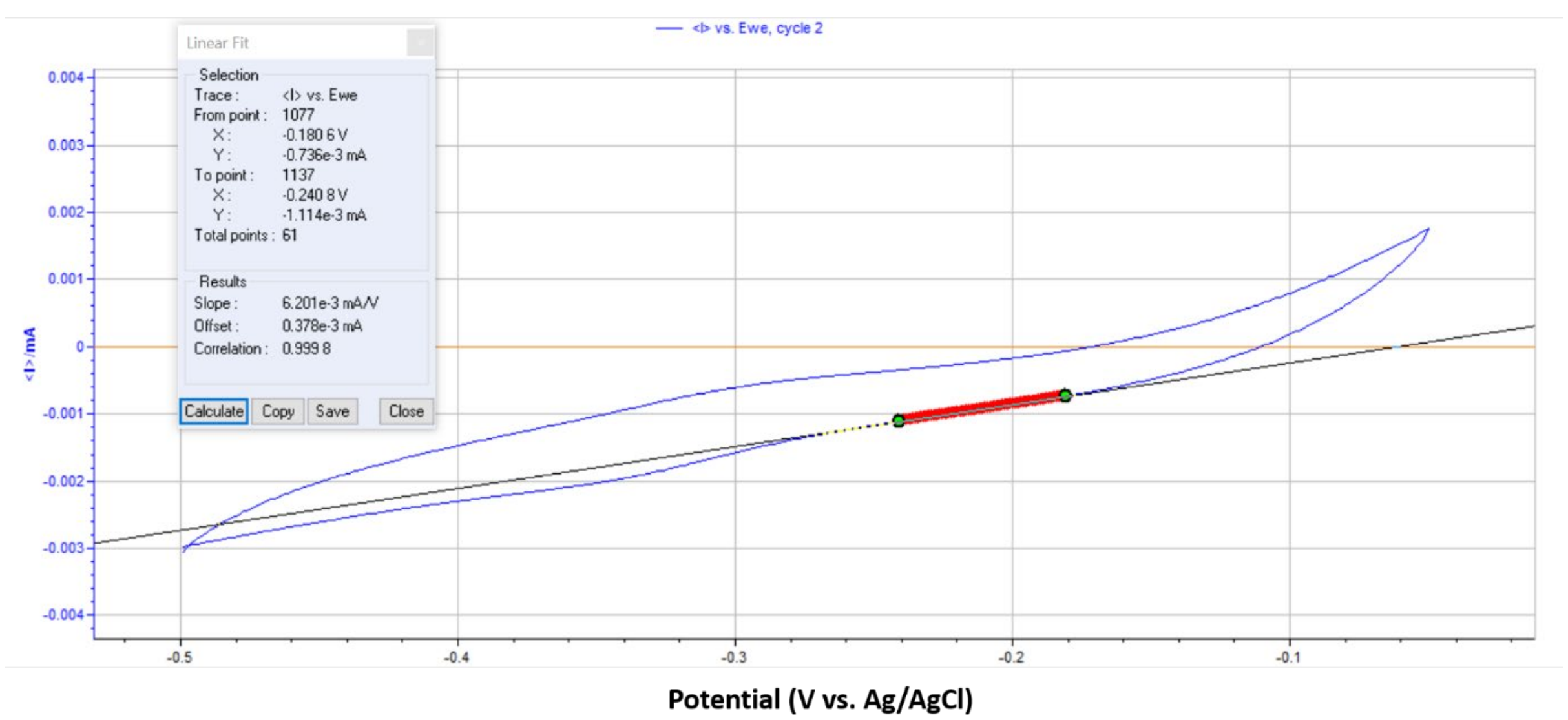

**e 5mV/s_Cycle 3 linear fitting**

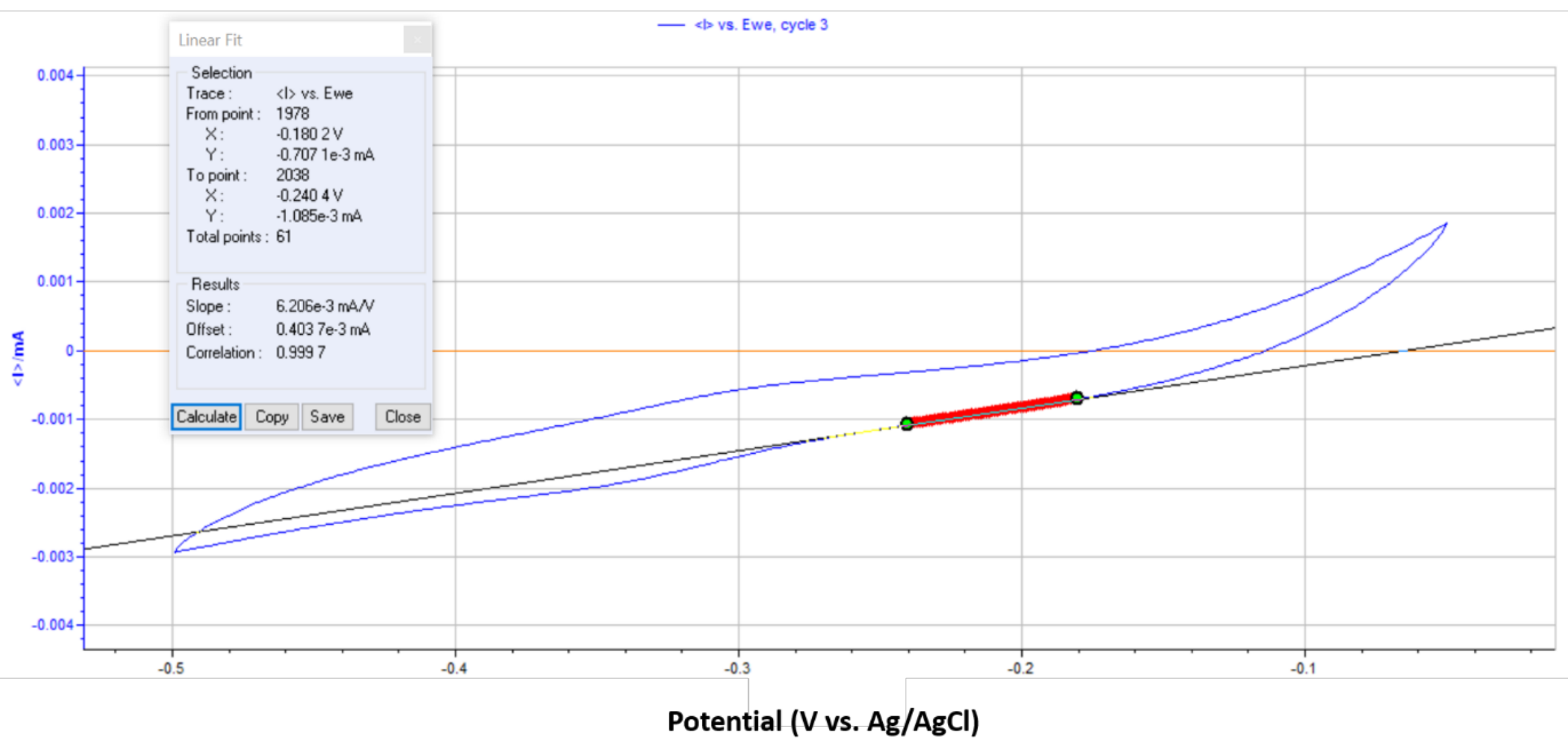

**Supplementary Figure 5. Tangent line drawings and schematics of how to extract Faradaic charges from cyclic voltammetry plots.** (a) Here we show the processing of the raw data of Cycle 1 at 5 mV/s as an example, where $\mathbf{i_1}$ and $\mathbf{i_2}$ represent the quenching onset and finishing point currents. $\mathbf{i_{NF}}$ denotes non-Faradaic current, and $\mathbf{i_F}$ denotes the Faradaic counterpart. (b)-(e) Linear fittings for CV data in 4 scenarios with results (slope and intercept) in the inset text boxes.

## M9. Method for developing working curve of fluorescence intensity vs. percentage of reduced functional groups in scattered RAC particles

Here we show the processing the data of Cycle 1 at 5 mV/s as an example. **In Supplementary Figure 6a**, green dashed lines denote the starting and ending of quenching in Cycle 1, with intercepts 36.3 s and 66.0 s on the time axis respectively. And we pick this time interval, overlay the Intensity vs. Time data with Faradaic Charge/Number of Functional Groups vs. Time data extracted from **M8**, so that a doubleY-axes plot is obtained (**Supplementary Figure 6b**). And after the two Y-axes values are correlated, we obtain a working curve(**Supplementary Figure 6c**).

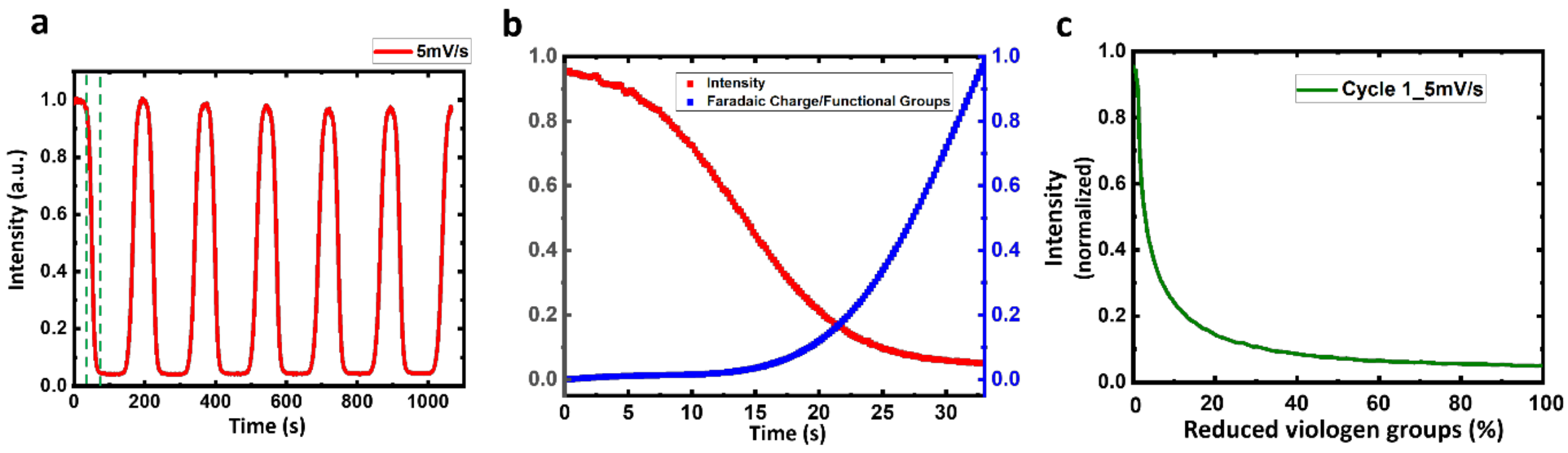

**Supplementary Figure 6. Process of developing working curve for Cycle 1 at 5 mV/s.** All the other working curves presented in **Fig. 1g** and **Supplementary Figure 16** are developed using the same workflow.

## M10. Fluorometry and photoluminescence (PL) measurements

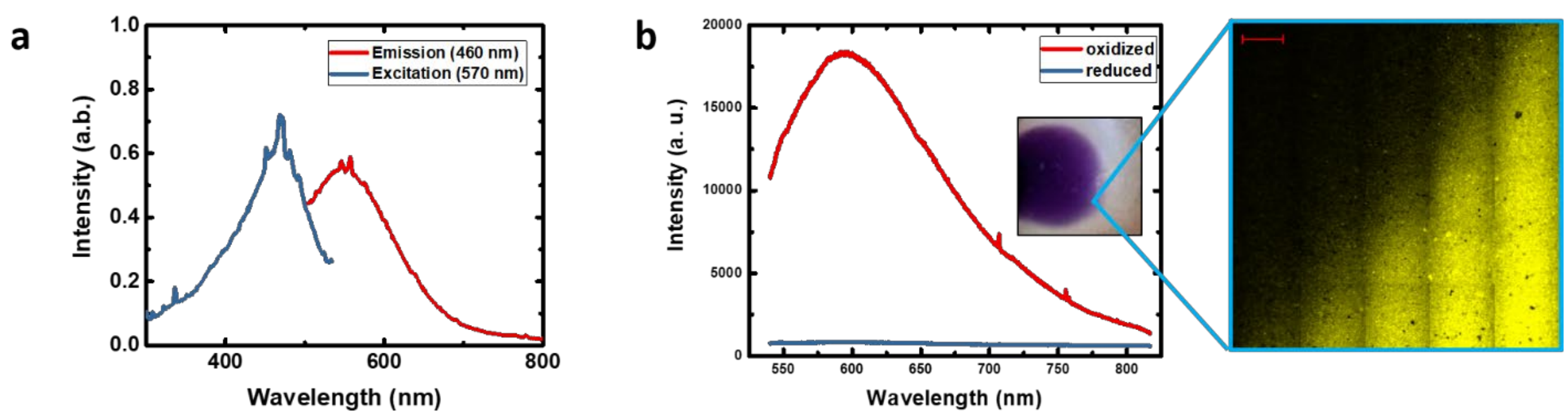

**Supplementary Figure 7. Photophysical properties of RAC and electrochromism of the bulk material.** (a) Emission and excitation scan data obtained from fluorometer of an oxidized RAC dispersion (10mM viologen groups). (b) Photoluminescence (PL) emission of oxidized and reduced RAC. Inset shows electrochromism of the bulk material, with violet part being reduced RAC (by zinc powder) and brighter orange-ish part being oxidized ones. On the right is the zoomed-in fluorescence imaging data. Scale bar is 20 μm. Note, stripes are an artifact due to inhomogeneity of the optical field.

## M11. Rough calculation showing chemical quenching ability of free-flowing reduced ethyl-viologen molecules ($EV^{+\cdot}$) in a RAC dispersion containing 10 mM oxidized viologen groups.

After an initial addition of $8.19\times10^{-5}$ M reduced ethyl-viologens to the RAC dispersion (10 mM ethyl-viologen groups), PL emission intensity around 570 nm decreased (**Supplementary Figure 8**) from 1799 (a.u.) to 632 (a.u.), which translates into one $EV^{+\cdot}$ molecule approximately quenching 10 mM $\times$ (1799-632) $\div$ 1799 $\div$ ($8.19\times10^{-5}$ M) $\approx$ 79 pendant groups in the RAC dispersion.

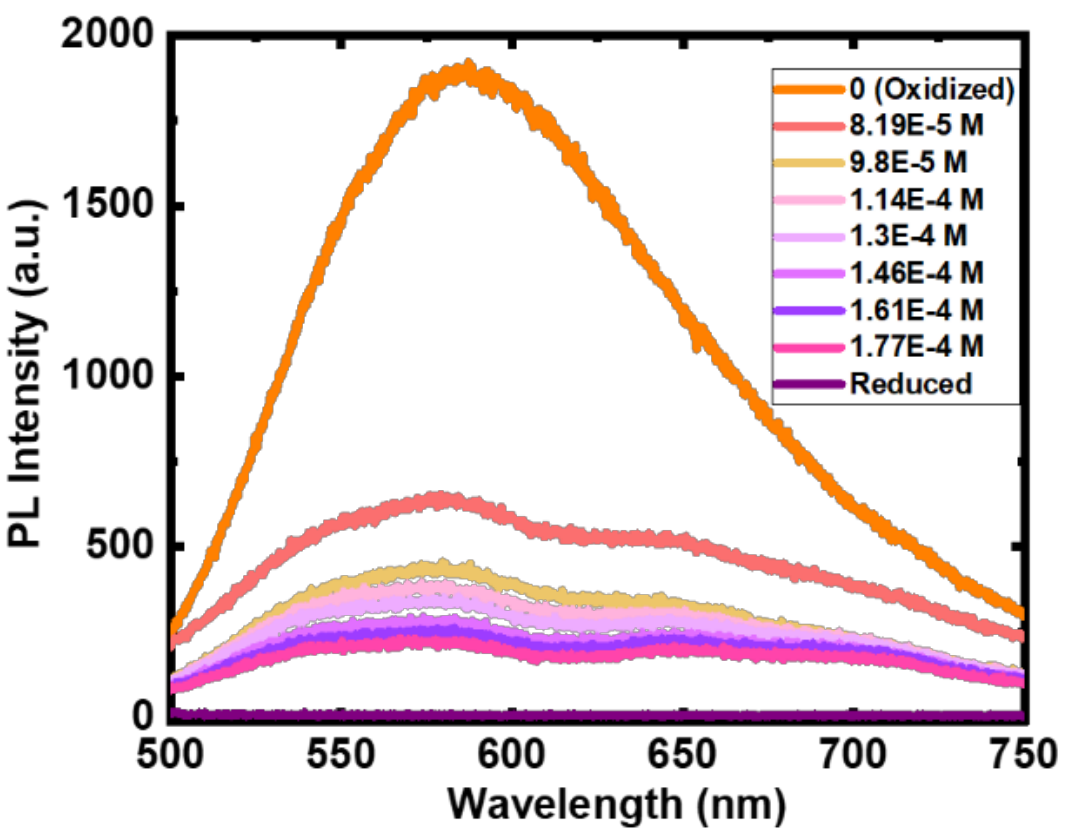

**Supplementary Figure 8. Photoluminescence emission data of oxidized RAC dispersion (10mM viologen groups) with stepwise addition of reduced ethyl-viologen small molecule ($EV^{+\cdot}$).**

### M12. PL measurement and time-resolved PL measurement procedures

The photoluminescence spectra are taken at room temperature, using a NKT super continuum laser (SuperK Extreme), filtered to 488 nm with a bandwidth of 5 nm and total power average on the sample of 2 mW. The laser is focused onto the sample by a parabolic mirror, which is also used to collect the luminescence from the sample. The collected luminescence is directed through a long -pass filter with a cut-off wavelength of 530 nm, to eliminate the scattered laser light from the sample. For the spectral measurements, the PL light is directed to an astigmatism-corrected Princeton Instruments (Spectra Pro 500i) 500 mm focal length spectrometer, equipped with a CCD camera. The time-correlated single photon counting measurements are performed using a Si single-photon counting avalanche photodiode and a Becker&Hickl TCSPC module (SPC-130).

### M13. Bulk electrolysis (reduction) of RACs

Electrolysis was performed on a CHI760 potentiostat and in an $O_2$ and moisture free environment inside of an Ar-filled drybox. A three-compartment cell was used, with the working electrode in the center compartment, and the two lateral compartments occupied by a carbon felt as a counter electrode, and a nonaqueous $Ag/Ag^+$ reference electrode (0.1 M $AgNO_3$, MeCN). A carbon felt on Pt wire was used as the working electrode and held at a constant overpotential while stirring. Current and charge response over time are recorded. For reduction, the potential is held −150 mV from $E_{1/2}$. UME (ultramicroelectrode) voltammograms using 12.5 µm radius Pt UMEs are obtained before and after electrolysis to track steady state limiting currents, confirming change of oxidation state.

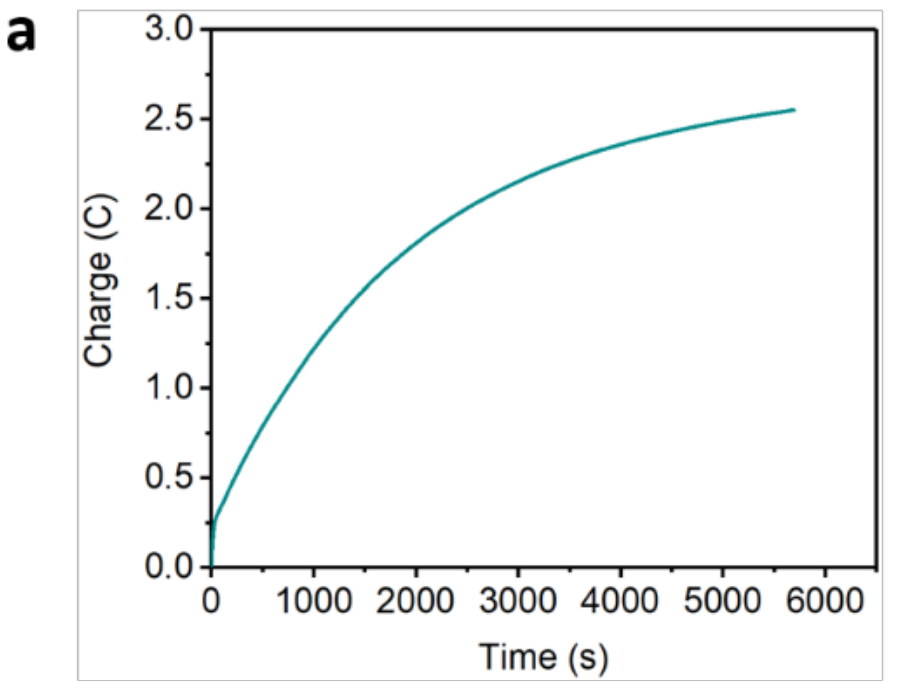

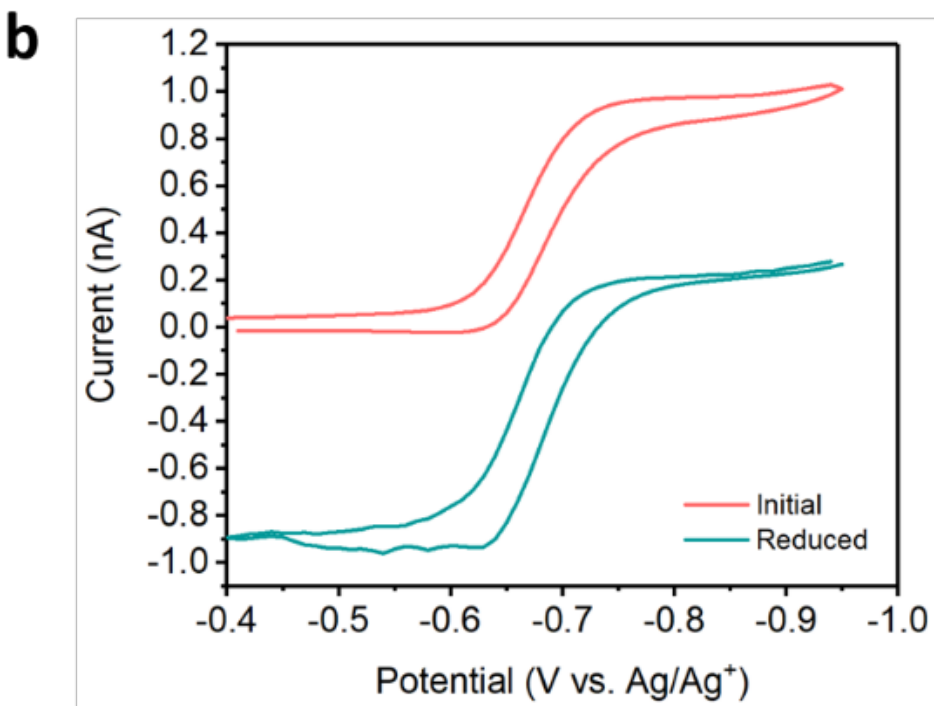

**Supplementary Figure 9. Bulk electrolysis of RAC.** (a) Charge vs. time relationship for electrolyzing 6.3 mM RAC (viologen groups) in 0.1 M $LiBF_4$ MeCN charged at -0.85 V vs. $Ag/Ag^+$. (b) UME-CVs show full conversion of RAC to the reduced state.

### M14. UV-Vis measurements of reduced RAC

UV-Vis absorption spectra are recorded on an HP8452A Diode-Array Spectrophotometer using a 1cm pathlength quartz cuvette. Dispersions of reduced RAC are diluted to 100 µM in acetonitrile and spectra are collected.

### M15. Propagation front tracking of RAC monolayer scenario

Fluorescence front in a colloidal monolayer was tracked by a home-built MATLAB code suite. Shown in **Supplementary Figure 10**, the electrode boundary position was first tracked by turning on the transmission so that electrode will block the light and show as dark. The fluorescence front inside the colloidal monolayer was tracked by the intensity difference between colloids that are reduced (dark) and oxidized (bright). Details of the tracking algorithm are given below.

Electrode boundary was first tracked from the optical images by scanning over the intensity line-by-line. From the optical image captured before reduction happens, a Gaussian filter (built-in MATALB function: imgaussfilt.m) with standard deviation 5 was applied to smooth image. For each horizontal line along $x$ axis (blue line), the fluorescence intensity profile is averaged by neighboring 5 pixels (red dots). The averaged intensity was then fitted by a tangent-hyperbolic[5] function: $f(x) = A + B \times \tanh((x - x_0)/w)$, in which $A$, $B$, $x_0$ and $w$ were four independent fitting parameters and $x_0$ was determined as the electrode boundary position for one fixed y coordinate (blue diamond). By applying this method to each y coordinate, the electrode boundary profile was extracted showing as red line and the approximate boundary was calculated from the linear regression of the boundary profile (green dotted line).

Fluorescence wave front beyond electrode in RAC monolayer was tracked following a similar algorithm used to map the electrode profile stated above. Starting from the original image, a Gaussian filter with standard deviation 3 was applied and the positions of the fluorescence front is extracted from this image by scanning the intensity profile of each horizontal line. From the intensity profile of a line scan, the intensity was fitted by the same tangent-hyperbolic function: $f(x) = A + B \times \tanh((x - x_0)/w)$ and $x_0 + w$ was determined as the position of fluorescence front for one fixed y coordinate (blue diamond). It is noteworthy that here, the fluorescence front was selected at the very edge when the fluorescence intensity begins to decrease. By applying this method to each y coordinate, the fluorescence front profile was extracted showing as red line.

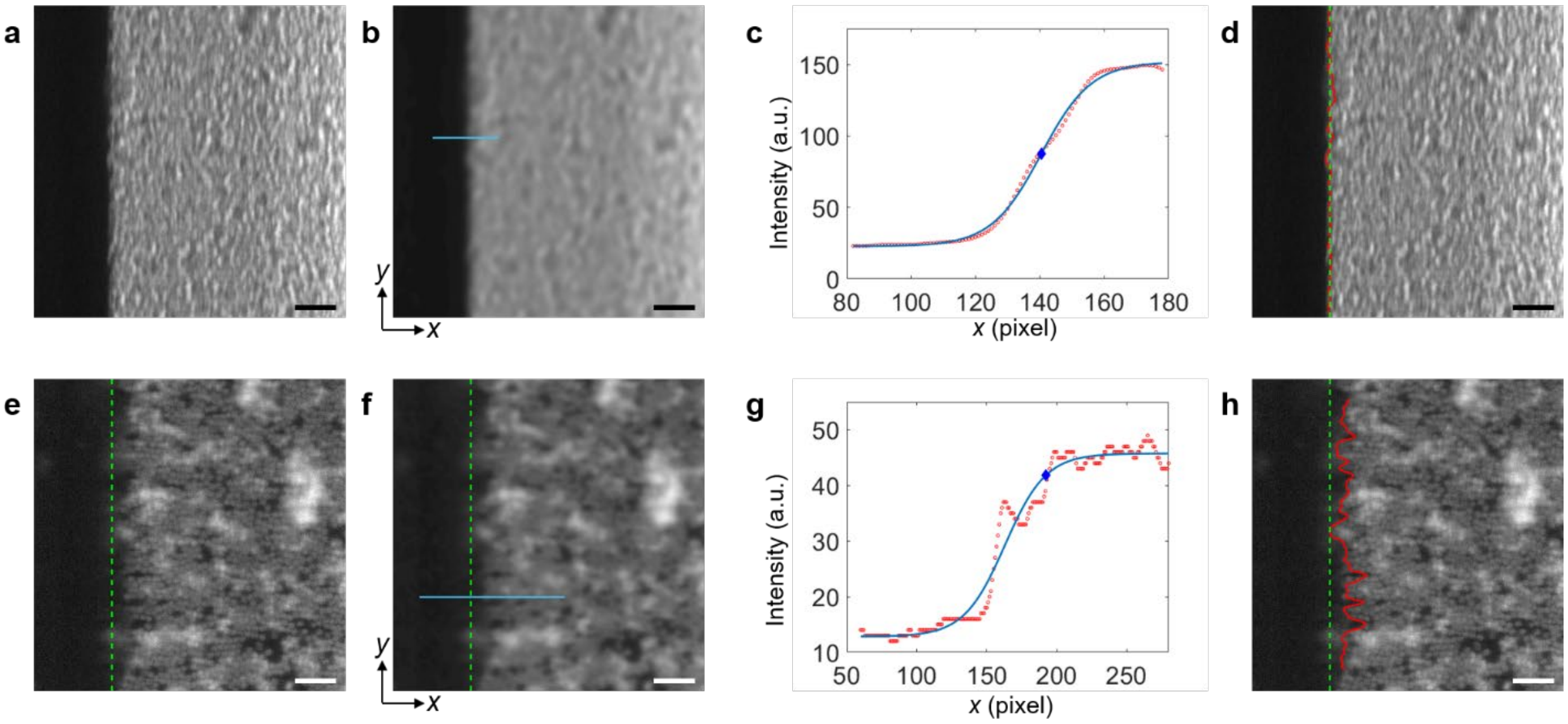

**Supplementary Figure 10. Automatic boundary detection of the electrode periphery and fluorescence wave front for RAC monolayer.** (a) An optical image of RAC monolayer before reduction with transmitted light turned on. The black region is the electrode. (b) The filtered image of (a) for boundary detection. The blue line indicates one position of horizontal intensity scan. (c) Intensity profile along the blue line in (b) after smoothing over neighboring 5 pixels (red dots). Blue line is fitted line and the blue diamond indicates the position of the electrode periphery. (d) The original image overlaid with the exact (red solid) and approximate (green dashed) boundaries. (e) A fluorescence image of RAC film after 24.6 minutes of reduction. The green dotted line is the approximate boundary from (d). (f) The filtered image of (e) for boundary detection. The blue line indicates one position of horizontal intensity scan. (g) Intensity profile along the blue line in (f) (red dots). Blue line is the fitted line and the blue diamond indicates the position of the fluorescence front. (h) The original image overlaid with the fluorescence front identified. Scale bars: 10 µm and single pixel size is 0.16 µm.

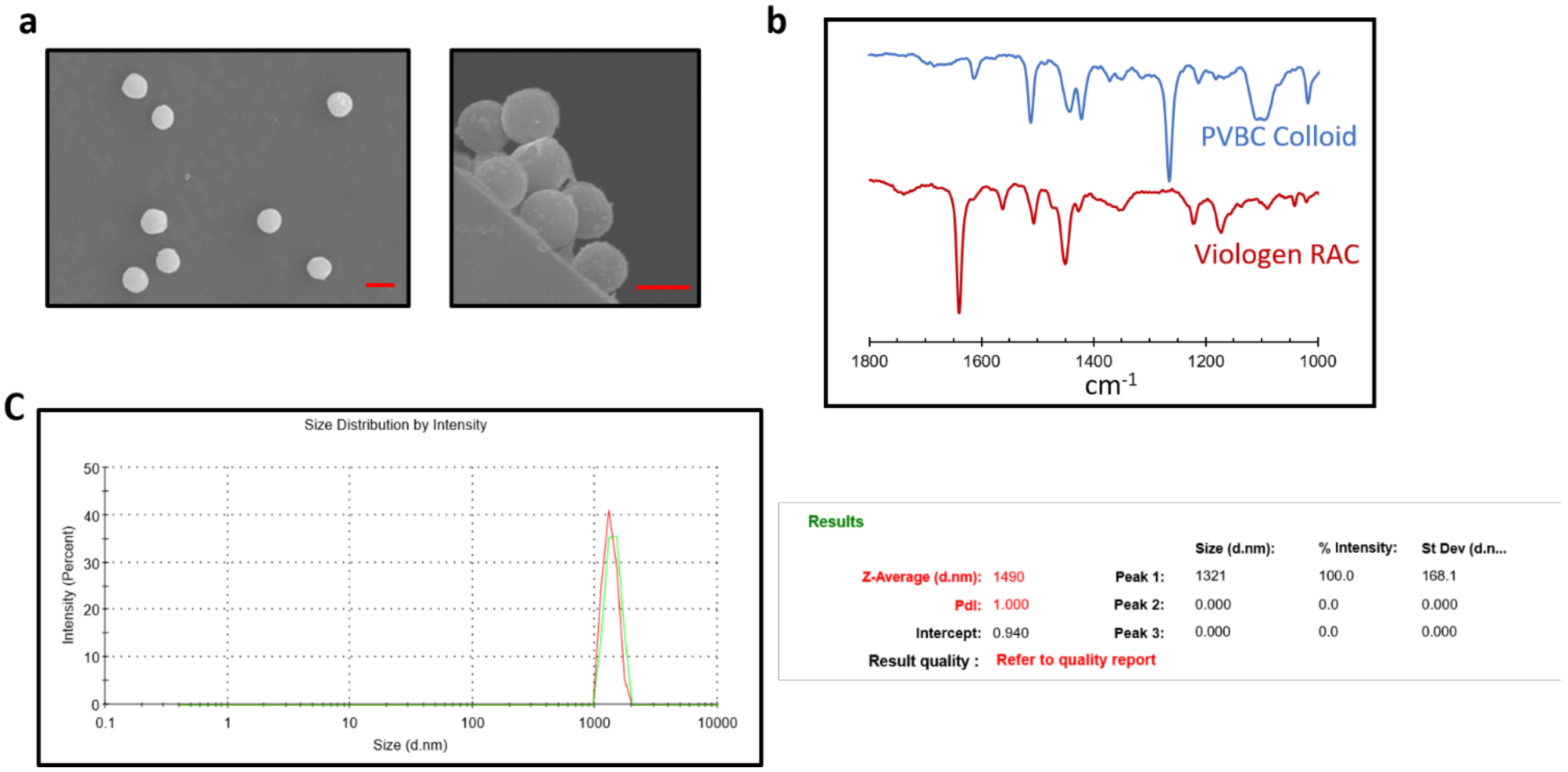

**Supplementary Figure 11. Basic Characterizations of RAC.** (a) Dry state Scanning Electron Microscopy (SEM) sizing with diameter is 950±32 nm (averaged over 50 particles). Scale bar 2 µm (both left and right). (b) ATR-IR spectra overlaid of pristine crosslinked poly (vinylbenzyl chloride) (xPVBC) colloid and ethyl-viologen derivatized ones (RAC), showing that C-Cl bond stretching (1240 $cm^{-1}$) dampened after viologen substitution[1] as well as C-cationic $N^+$ vibrational modes appearance (1640 $cm^{-1}$)[1]. (c) Dynamic light scattering (DLS) measurements of 0.1g/L oxidized RAC size in acetonitrile, 1321±160 nm. (Also, for elemental analysis see previous work[1].)

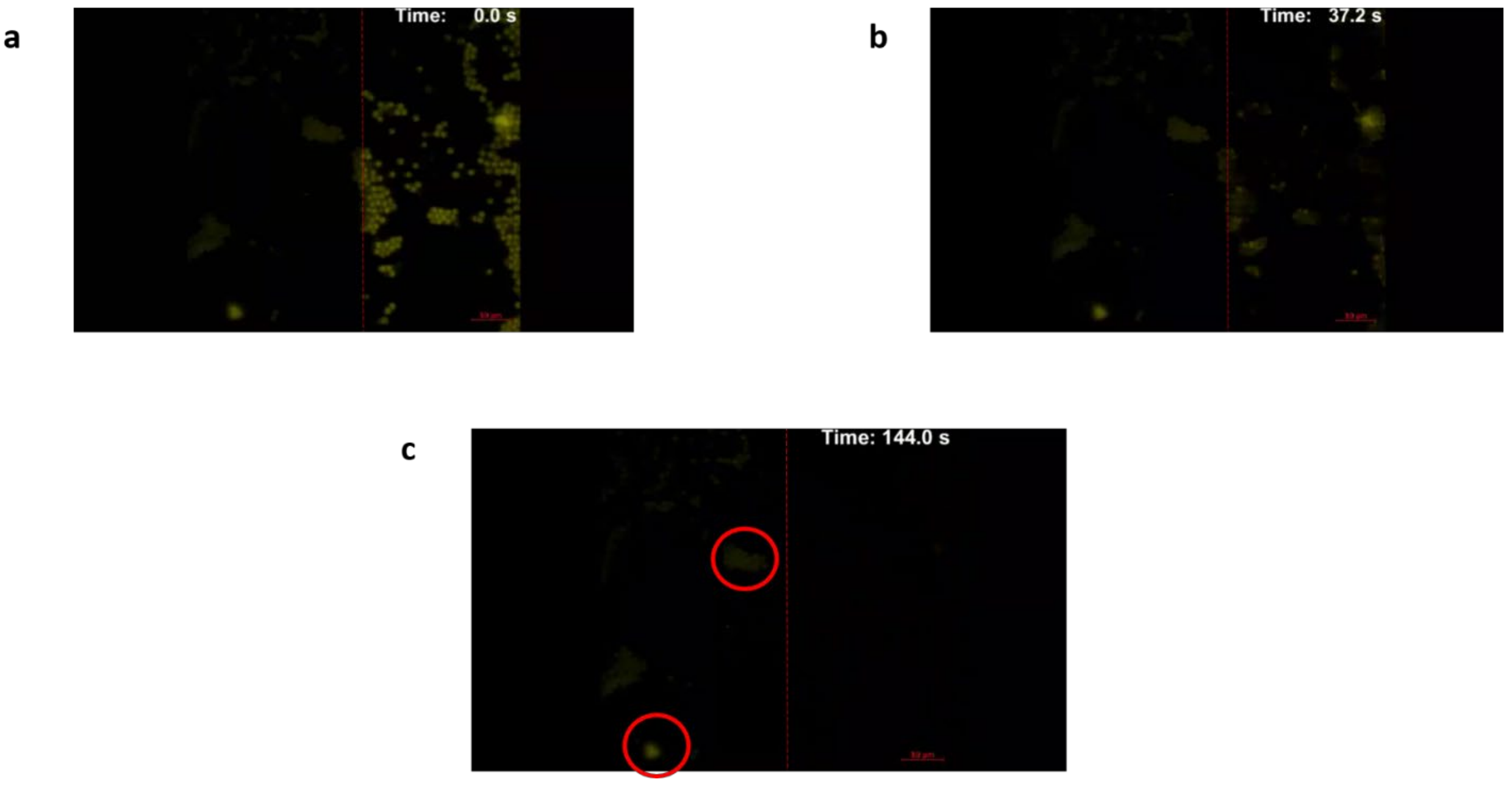

**Supplementary Figure 12. Charge transport among non-percolating RAC clusters (CV: 20 mV/s).** (a,b,c): Time-lapsed snapshots of fluorescence quenching in a non-percolated case. Right side of the image is the electrode and left side is the insulating gap (glass slide as substrates), with red dashed line denoting electrode boundary. From (c) it is seen clearly that those isolated RAC clusters did not lose fluorescence emission at a sufficiently long time, indicating electron transport can only happen upon physical contact. Scale bars: 10 μm.

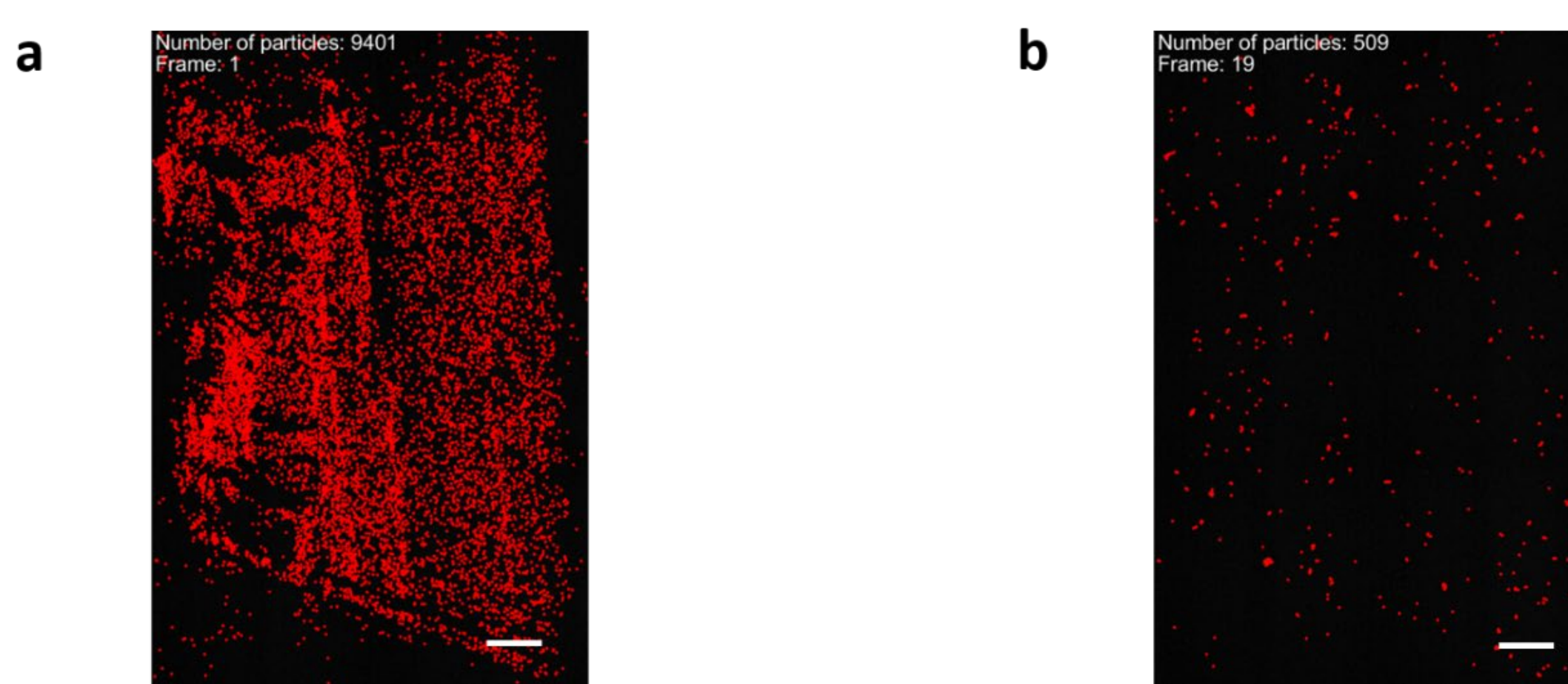

**Supplementary Figure 13. Low-magnification (5X) large-area snapshots before (a) and after (b) reducing RAC on the entire working electrode (sweep rate: 10 mV/s).** (a) Particle counting result: 9401. (b) Particle counting result: 509. Scale bars: 200 μm.

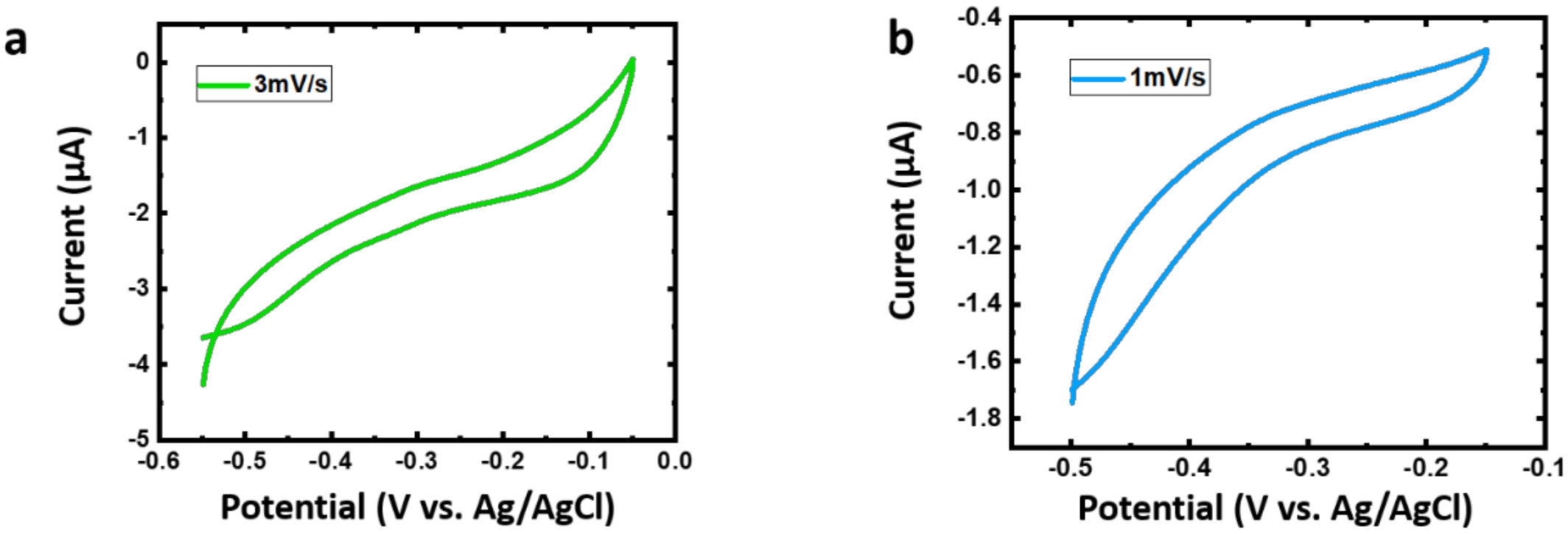

**Supplementary Figure 14. Cyclic Voltammetry run on discrete RAC pinned onto the working electrode.** (a) sweep rate: 3 mV/s. (b) sweep rate: 1 mV/s. Cast RAC concentration and method are the same as shown in **Supplementary Figure 3**.

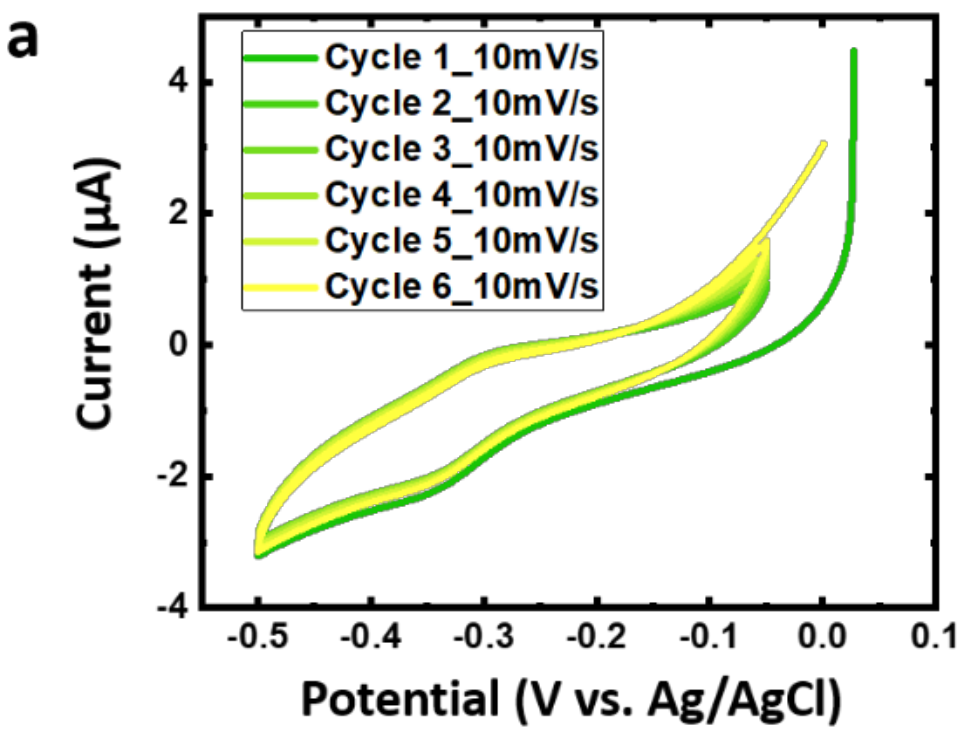

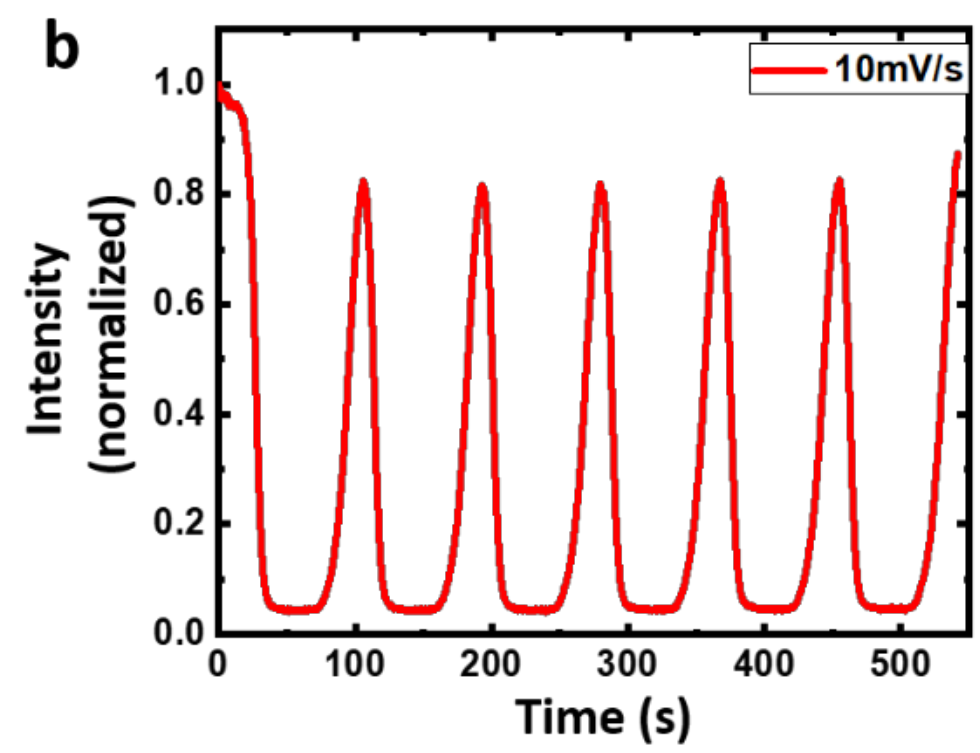

**Supplementary Figure 15. Data used to develop working curve at 10mV/s in Fig. 1g.** (a) 6-round CV cycles run on 34,731 colloids. (b) Ensemble average fluorescence intensity vs. time of the 166 particles in the field of view (**Supplementary Figure 18**). After the 1st round, ensemble intensity did not return to the original intensity because this scan rate does not provide enough time to fullyelectrolyze the RAC.

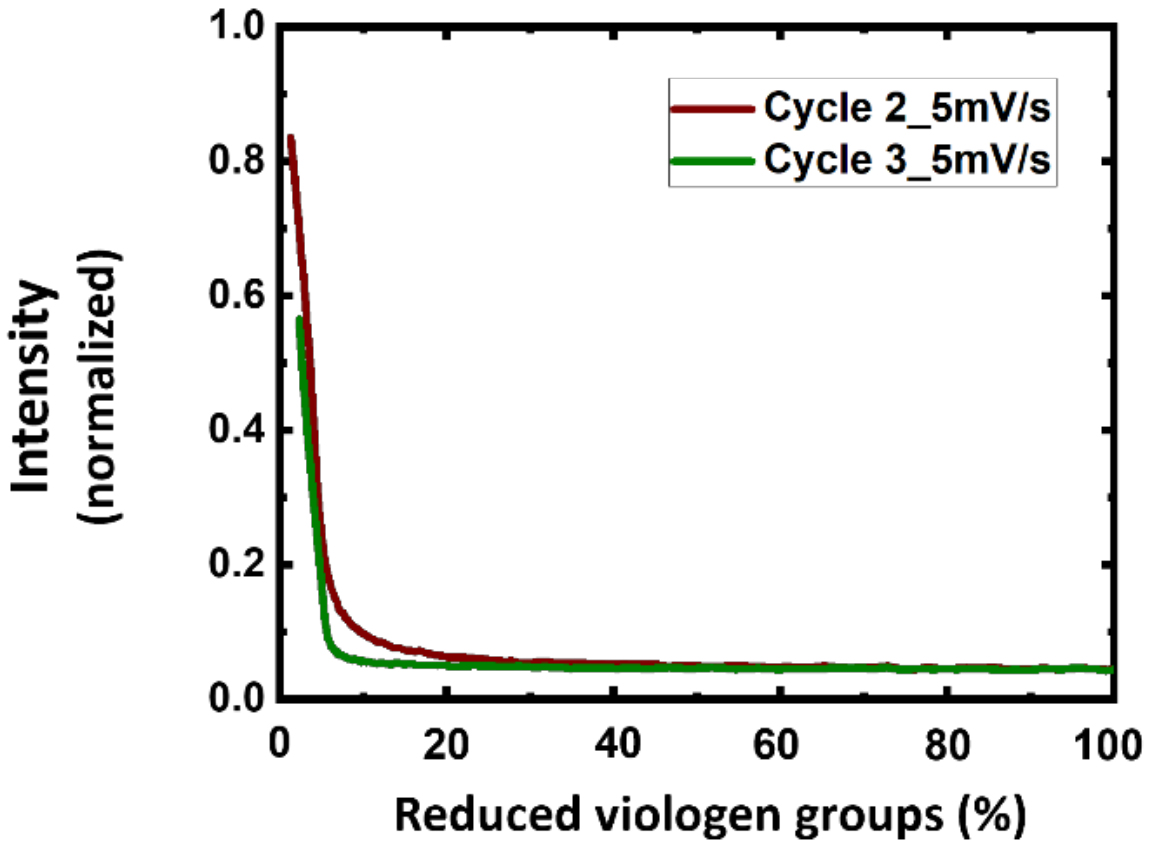

**Supplementary Figure 16. Fluorescence intensity (normalized) vs. percentages of reduced pendant groups in RAC extracted from Cycle 2 and 3 at 5mV/s sweep rate , developed with the method shown in Supplementary Figure 4.**

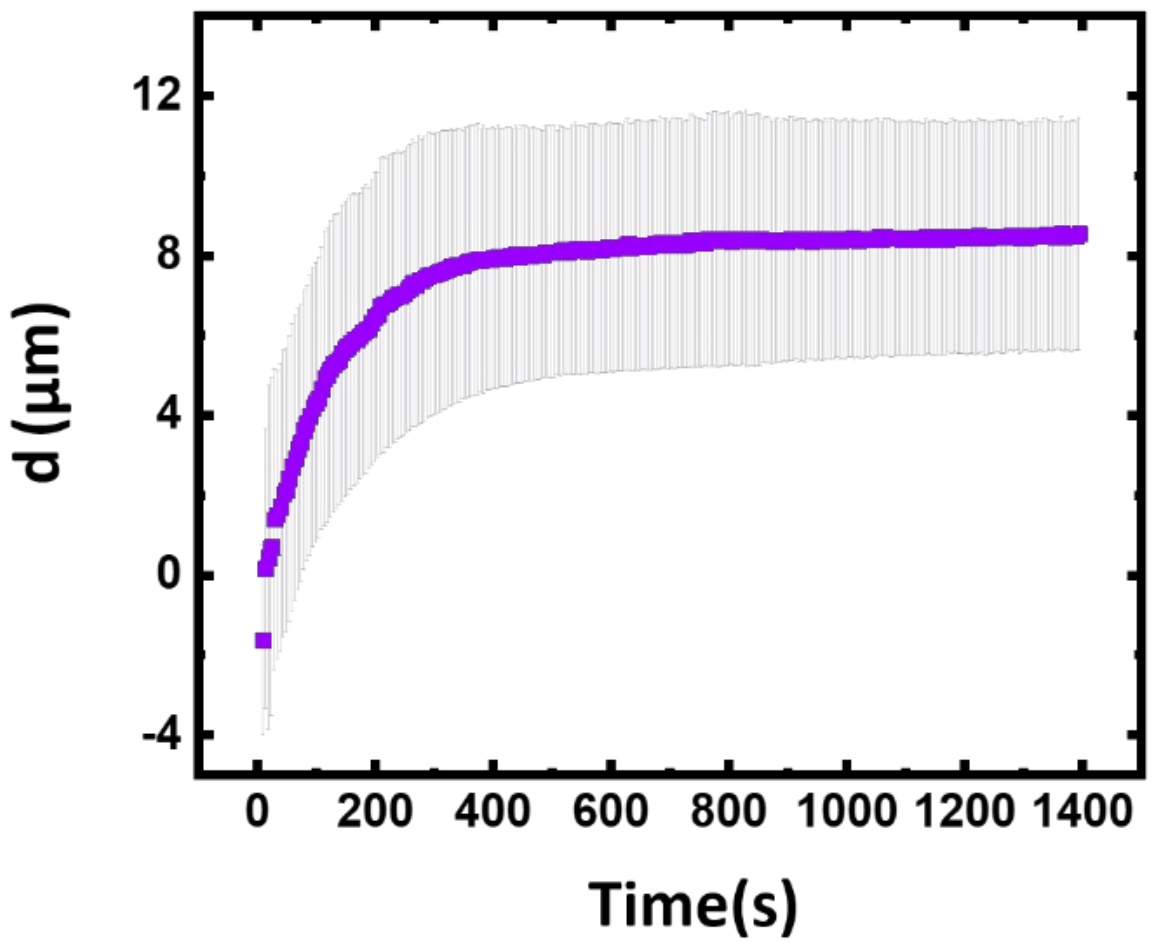

**Supplementary Figure 17. Lateral distance(d) between fluorescence wave front and Pt/glass boundary over time, for the reduction part in Fig 3c.** Shaded area denotes the error bar, attributed to tracking multiple horizontal lines orthogonal to the fluorescence wave front.

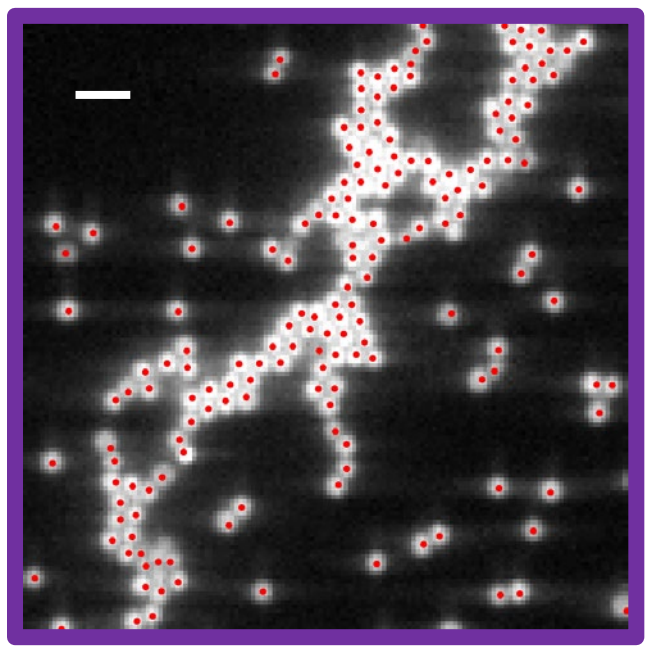

**Supplementary Figure 18. Particle-counting in the field of view (166 particles in total, scale bar is 10 μm) corresponding to data in Fig.1e.**

## Supplementary Movies

**Movie S1: Synchronized data of RAC monolayer fluorescence switching on the working electrode at 1mV/s sweep rate**

**Movie S2: Synchronized data for extracting the working curve at 5mV/s sweep rate**

**Movie S3: Synchronized data for extracting the working curve at 10mV/s sweep rate**

**Movie S4-7: Synchronized data of RAC monolayer lateral charge transport (Cycle 1-4)**

**Movie S8: Demonstration of no charge transport in non-percolating RAC clusters on the insulating glass substrate (cyclic voltammetry at 20mV/s sweep rate)**

**Movie S9: Low-magnification large-area fluorescence switching imaging on the entire working electrode at 10mV/s sweep rate**